\documentclass{aa} 

\usepackage{graphicx}
\usepackage{txfonts}
\usepackage{lipsum}
\usepackage{subcaption}        
                              
\usepackage{lscape}            
                               
\usepackage{placeins}

\begin{document}

\title{Only obscured yet luminous active galactic nuclei are closely associated with galaxy mergers: Direct observational evidence from type 2 active galactic nuclei}
\titlerunning{Only obscured yet luminous AGNs are closely associated with galaxy mergers}

\author{
Yongmin Yoon\inst{1}\thanks{yyoon@knu.ac.kr}
\and
Yongjung Kim\inst{2,3}\fnmsep\thanks{Corresponding author; yongjungkim@sejong.ac.kr}
\and
Dohyeong Kim\inst{4}\fnmsep\thanks{Corresponding author; dohyeongkim.ast@gmail.com}
\and
Jaejun Cho\inst{1}
\and
Woowon Byun\inst{5}
}
\institute{
Department of Astronomy and Atmospheric Sciences, College of Natural Sciences, Kyungpook National University, Daegu 41566, Republic of Korea
\and
School of Liberal Studies, Sejong University, 209 Neungdong-ro, Gwangjin-Gu, Seoul 05006, Republic of Korea
\and
Department of Physics and Astronomy, Sejong University, 209 Neungdong-ro, Gwangjin-Gu, Seoul 05006, Republic of Korea
\and
Department of Earth Sciences, Pusan National University, Busan 46241, Republic of Korea
\and
Korea Astronomy and Space Science Institute (KASI), 776 Daedeokdae-ro, Yuseong-gu, Daejeon 34055, Republic of Korea
}
\authorrunning{Yoon et al.}

\abstract{To establish a more comprehensive understanding of the connection between galaxy mergers and active galactic nuclei (AGNs), it is essential to disentangle the contributions of intrinsic AGN luminosity and dust extinction to the merger--AGN connection. Since tidal features identified in deep images serve as direct evidence of recent mergers, we studied the fraction of AGN hosts with tidal features ($f_T$) for a large sample of 748 type 2 AGNs at $z<0.063$. Specifically, we examined $f_T$ as a function of $E(B-V)$, derived from the Balmer decrement, and the internal-extinction-corrected luminosity of the [O\,{\footnotesize III}] $\lambda$5007 emission line ($L_\mathrm{[O\,{\footnotesize III}]}$), which is a proxy for bolometric AGN luminosity. Our main finding is that $f_T$ is only significantly higher for AGNs that are simultaneously luminous and heavily dust-obscured. Specifically, AGNs with $\log L_\mathrm{[O\,{\footnotesize III}]}\gtrsim41.5$ and $E(B-V)\gtrsim0.7$ exhibit a high $f_T$ of $\sim0.7$. In contrast, AGNs with either low luminosity ($\log L_\mathrm{[O\,{\footnotesize III}]}\lesssim41.0$) or low dust obscuration ($E(B-V)\lesssim0.3$) show a low $f_T$ of $\lesssim0.2$. This trend suggests that galaxy mergers preferentially trigger AGNs that are simultaneously luminous and dust-obscured, whereas AGNs that are either luminous but unobscured or dust-obscured but less luminous are not strongly associated with merger-driven triggering. Based on several assumptions, our result can also be interpreted, despite certain caveats, within the framework of a merger-initiated evolution model for AGNs, suggesting that AGNs that are both obscured and luminous are temporally closer to merger events than those with lower luminosities and less dust obscuration.
}

\keywords{galaxies: active -- galaxies: interactions -- quasars: general -- quasars: supermassive black holes}

\maketitle
\nolinenumbers

\section{Introduction}\label{sec:intro}

Active galactic nuclei (AGNs) are unique cosmic laboratories for investigating fundamental properties and processes of the universe. For instance, AGNs offer valuable clues about the evolutionary paths of galaxies, as they are expected to regulate star formation and galaxy growth  \citep{DiMatteo2005,Bower2006,Croton2006,Hopkins2008,Arjona2024} via energetic outflows such as winds and jets \citep{Santoro2020,Torrey2020,Laha2021}. This coevolution between AGNs and their host galaxies is evident in tight correlations observed between the masses of supermassive black holes (SMBHs) and the properties of their host galaxies \citep{DiMatteo2005,Kormendy2013,McConnell2013}. 

Moreover, AGNs provide ideal environments for studying physics under extreme electromagnetic and gravitational fields \citep{Dovciak2004,Jovanovic2008,Blandford2019}. In addition, certain AGNs are extremely luminous, enabling their detection at extreme cosmic distances and providing a window into the early Universe \citep{Kim2015a,Kim2019,Kim2020,Jiang2016,Wang2021,Sacchi2022,Christensen2023}. Finally, AGN radiation, especially from luminous ones, can be used as a backlight for absorption studies of intergalactic and circumgalactic media, through which the properties of diffuse gas in and around galaxies can be probed \citep{Tumlinson2017,Dalton2022}.

Active galactic nuclei can be triggered through a variety of mechanisms. Internal processes, such as the influence of galactic bars or disk instabilities, can funnel material toward the central SMBHs, triggering AGN activity \citep{Shlosman1989,Crenshaw2003,Ohta2007,Hirschmann2012,Menci2014}. Additionally, the inflow of cold gas from the hot halos of galaxy clusters can serve as a source of fuel \citep{Li2014,Tremblay2016}. Furthermore, external environmental effects, such as ram-pressure experienced by galaxies moving rapidly through dense cluster environments, can contribute to triggering AGN activity \citep{Poggianti2017}. 

Galaxy mergers are also recognized as triggering mechanisms for AGNs. During a merger, gravitational interactions generate tidal torques that remove angular momentum from the gas. This enables the gas to fall toward the galactic core, where it can accrete onto the SMBH and trigger AGN activity \citep{DiMatteo2005,Springel2005,Hopkins2008,Capelo2015}.

Numerous observational studies have explored the potential connection between galaxy mergers and the triggering of AGN activity. A number of studies have found AGNs to be more prevalent in galaxies undergoing interactions or in post-merger phases, compared to inactive counterparts \citep{Carpineti2012,Cotini2013,Ellison2013,Hong2015,Marian2020,Araujo2023,Hernandez2023,Li2023,Comerford2024,Yoon2025}. This higher occurrence of mergers among AGN host galaxies highlights galaxy mergers as a plausible mechanism for triggering AGN activity. However, several studies have reported a lack of clear evidence for a merger-driven origin of AGN activity \citep{Gabor2009,Cisternas2011,Kocevski2012,Sabater2015,Mechtley2016,Villforth2014,Villforth2017,Marian2019,Shah2020,Zhao2022}.

Some studies suggest that galaxy mergers are more frequently associated with the triggering of high-luminosity AGNs, implying a  connection between merger events and AGN luminosity \citep{Treister2012,Ellison2013,Urbano2019,Hernandez2023,Pierce2023,Tang2023,Euclid2025,Yoon2025}. However, other studies have not observed this pattern \citep{Villforth2014,Marian2020,Steffen2023,Comerford2024}, suggesting that the relation between AGN luminosity and galaxy mergers remains inconclusive.

It is also known that not only high-luminosity AGNs, but also dust-obscured AGNs---whose internal extinction by dust in the host galaxy is high enough to significantly attenuate the intense radiation from the central engine---are commonly found in galaxies with signs of recent mergers \citep{Urrutia2008,Satyapal2014,Glikman2015,Kocevski2015,Weston2017,Goulding2018,Koss2018,Ellison2019,Secrest2020}. Many studies, based on both observations and simulations, interpret this trend in obscured AGNs, particularly those with high luminosities, within an evolutionary framework related to galaxy mergers \citep{Sanders1988,Cattaneo2005,Hopkins2008,Urrutia2008,Urruita2012,Banerji2012,Glikman2012,Glikman2015,Kim2015b,Kim2024a,Kim2024b,Ricci2017,KI2018}. This model begins with gas-rich mergers that trigger intense starbursts and fuel the central SMBHs. In the early phase of the merger, large amounts of dust and gas enshroud the nuclear region, blocking the radiation from the AGN. As the AGN evolves, AGN-driven feedback expels the obscuring material, transitioning the AGN into an unobscured phase. Therefore, according to this model, obscured AGNs may represent an early phase in the evolutionary sequence initiated by gas-rich mergers, implying that these AGNs are closely linked to galaxy mergers in both origin and temporal proximity to the merger event.

In many previous studies, dust-obscured AGNs have been selected based on mid-infrared colors, and this selection technique tends to preferentially identify intrinsically high-luminosity AGNs with bolometric luminosities above $\sim10^{45}$ erg s$^{-1}$ \citep{Cardamone2008,Weston2017,Donley2018,Goulding2018,LaMarca2024}.\footnote{Obscured AGNs tend to be intrinsically luminous \citep{Hopkins2006,Glikman2015,Zhuang2020,Kim2024b}.} This can bias the conclusions of previous studies on obscured AGNs and mergers, since the connection between mergers and AGN activity likely depends on both the degree of obscuration and AGN luminosity. Therefore, the merger--AGN connection should be examined quantitatively as a joint function of AGN luminosity and obscuration by disentangling the contributions of two key variables---dust extinction and intrinsic AGN luminosity---in order to obtain a more comprehensive understanding not only of the merger--AGN connection, but also of the evolutionary scenario of merger-triggered AGNs. In particular, examining low-luminosity AGNs with high obscuration and high-luminosity AGNs with low obscuration enables us to acquire a more complete view. 

For this purpose, we examined the fraction of AGN hosts exhibiting merger features in relation to AGN luminosity and the degree of obscuration, $E(B-V)$, derived from the Balmer decrement, using a large sample of 748 type 2 AGNs at $z<0.063$. Type 2 AGNs are an ideal sample for this study, as they allow for a quantitative investigation of a wide range of AGN luminosities and dust obscuration levels using a large dataset. 

Tidal features, which are remnants of stellar debris produced by galaxy mergers \citep{Quinn1984,Barnes1988,Hernquist1992,Feldmann2008}, represent evidence of recent merger events. Tidal features offer a way to overcome a key limitation in observational studies of galaxy mergers, that is, our inability to directly observe cosmic-scale processes over time since astronomical observations capture only a snapshot of the Universe. As a result, tidal features have been widely used in observational studies as reliable indicators of recent galaxy mergers \citep{Schweizer1992,Tal2009,Kaviraj2011,Sheen2012,Sheen2016,Duc2015,YL2020,Yoon2022,Yoon2023,Yoon2024a,Yoon2024b,Bilek2023}. Since these features are typically much fainter than the main structure of galaxies, deep images are required to detect them. In this study, we detected tidal features using deep-imaging data from the Dark Energy Spectroscopic Instrument (DESI) Legacy Imaging Survey \citep{Dey2019}.

Throughout this paper, we adopt the following cosmological parameters. We assume a Hubble constant of $H_0 = 70$ $\mathrm{km\,s}^{-1}\,\mathrm{Mpc}^{-1}$, a dark energy density parameter of $\Omega_{\Lambda} = 0.7$, and a matter density parameter of $\Omega_m = 0.3$.
\\

\begin{figure}
\includegraphics[width=\linewidth]{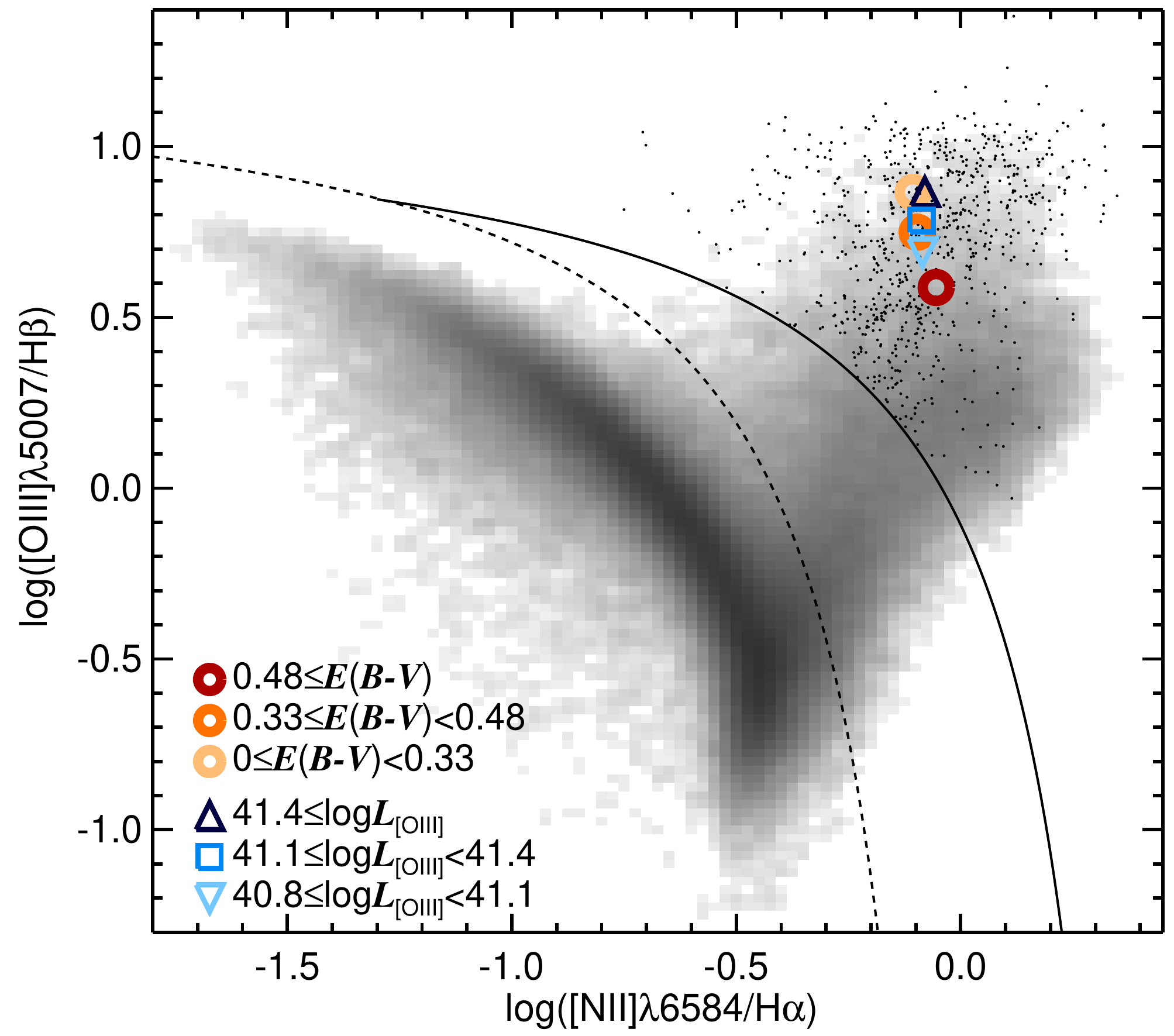}
\centering
\caption{Distribution of galaxies at $z<0.063$ in the BPT diagram, which classifies galaxies based on the emission-line flux ratios  [O\,{\footnotesize III}] $\lambda$5007/H$\beta$ and [N\,{\footnotesize II}] $\lambda$6584/H$\alpha$, is shown as a number-density plot in which darker regions indicate a higher density of data points. The dashed line indicates the demarcation defined by Equation (1) in \citet{Kauffmann2003}, below which galaxies are classified as star-forming. The solid line represents the classification boundary given by Equation (5) in \citet{Kewley2001}, above which galaxies are classified as AGNs. Galaxies located between the two lines are classified as composite, where both star-forming regions and AGNs contribute to the emission-line fluxes. Our 748 type 2 AGNs with $\log L_\mathrm{[O\,{\footnotesize III}]}\ge40.8$ are plotted as black circles. The colored symbols mark the median BPT positions of type 2 AGNs in bins of $E(B-V)$ and $L_\mathrm{[O\,{\footnotesize III}]}$ (an indicator of AGN luminosity), where each symbol is plotted at the median values of [N\,{\footnotesize II}] $\lambda$6584/H$\alpha$ and [O\,{\footnotesize III}] $\lambda$5007/H$\beta$ for the AGNs in that bin. Different symbols and colors denote different bins. $L_\mathrm{[O\,{\footnotesize III}]}$ is in units of erg s$^{-1}$.
\label{fig:bpt}}
\end{figure}

\begin{figure}
\includegraphics[width=\linewidth]{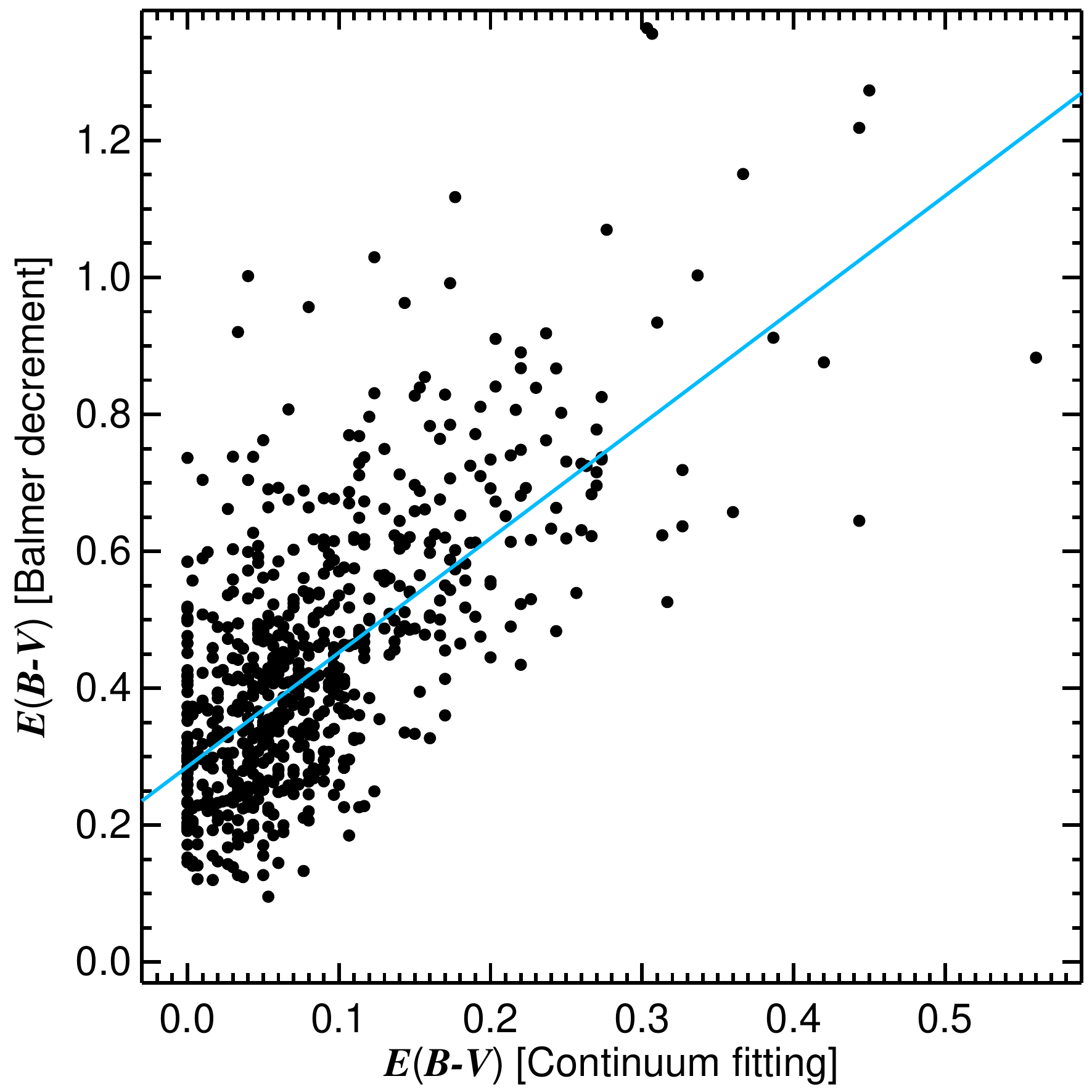}
\centering
\caption{Comparison between $E(B-V)$ derived from the Balmer decrement and that obtained from stellar continuum fitting for 662 type 2 AGNs. The blue line represents the best-fit relation determined using the least absolute deviation method ($y=1.67x+0.29$). The two parameters are well correlated, with a Pearson correlation coefficient of 0.67 and a mean of the absolute deviation of 0.1 from the best-fit relation. However, $E(B-V)$ derived from the Balmer decrement is, on average, 0.35 higher than that obtained from stellar continuum fitting.
\label{fig:ebv}}
\end{figure}

\begin{figure*}
\includegraphics[width=\linewidth]{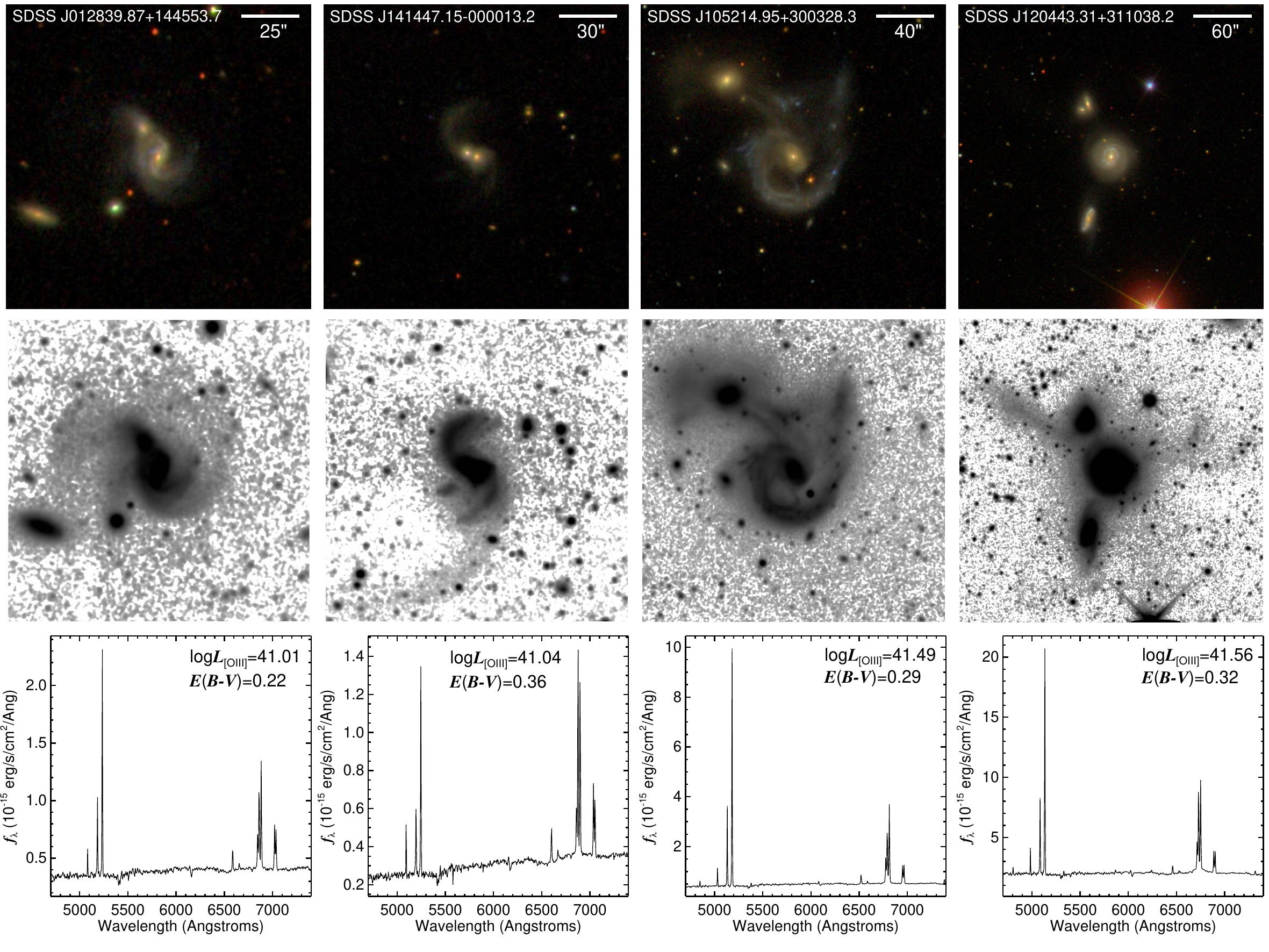}
\centering
\caption{Examples of type 2 AGN hosts with tidal features, which have low $E(B-V)<0.4$. The two left columns show AGNs with low $\log L_\mathrm{[O\,{\footnotesize III}]}<41.1$, while the two right columns display those with high $\log L_\mathrm{[O\,{\footnotesize III}]}>41.4$. First row: SDSS color images. The galaxy ID is displayed on each image, and a horizontal bar indicates the angular scale of the image. Second row: $r$-band deep images from the DESI Legacy Imaging Survey, displayed at the same angular scale as the color images in the first row. These images are shown in inverted grayscale after Gaussian smoothing with $\sigma=2$ pixels. Third row: Optical spectra of type 2 AGNs covering the observed wavelength range of 4700--7400\AA, where the H$\beta$, [O\,{\footnotesize III}], and H$\alpha$ emission lines are visible. The corresponding values of $\log L_\mathrm{[O\,{\footnotesize III}]}$ (erg s$^{-1}$) and $E(B-V)$ are also indicated.
\label{fig:ex_1}}
\end{figure*}

\begin{figure*}
\includegraphics[width=\linewidth]{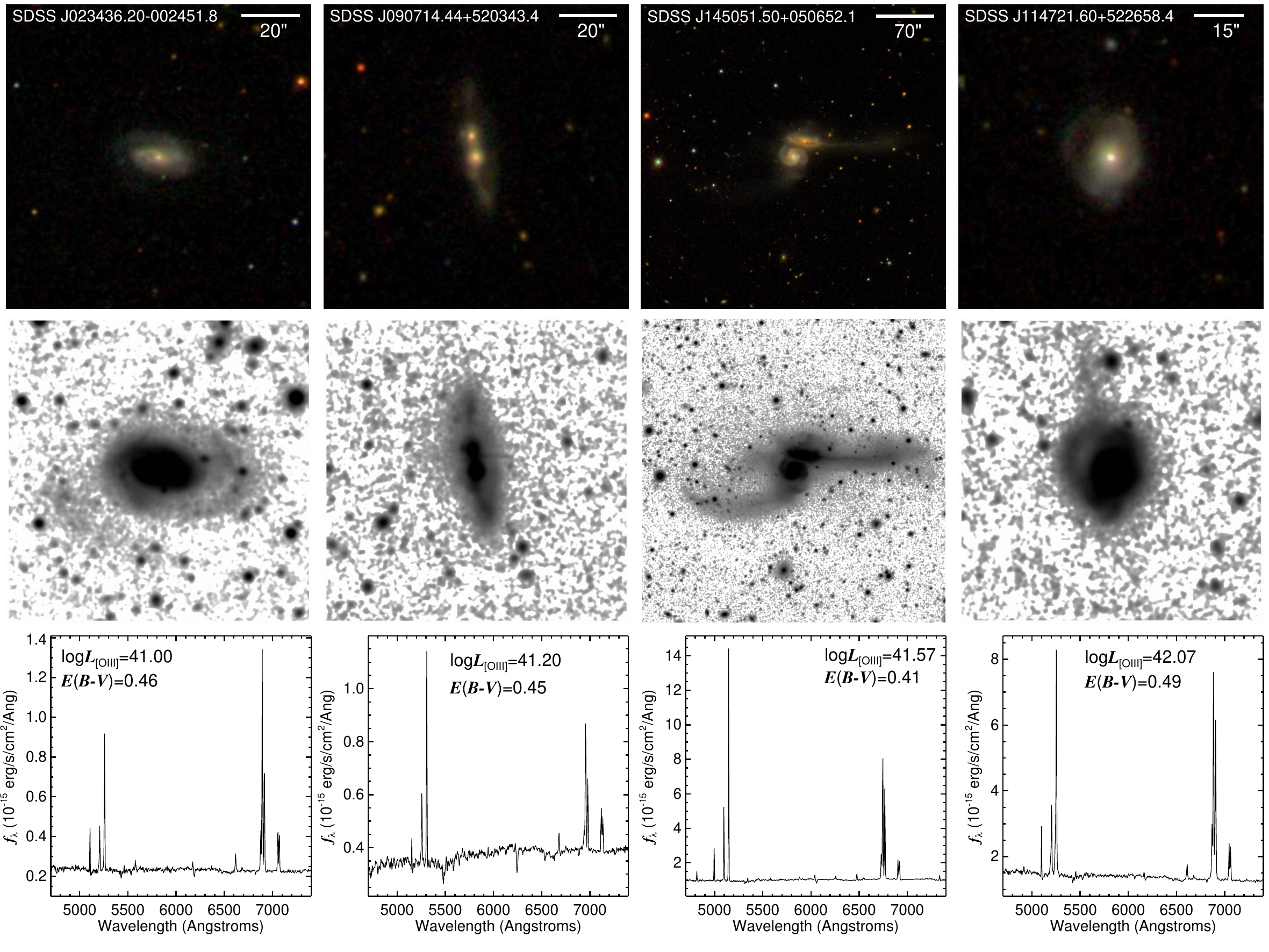}
\centering
\caption{Examples of type 2 AGN hosts with tidal features, which have intermediate $E(B-V)$ values between 0.4 and 0.5. The two left columns show AGNs with low $\log L_\mathrm{[O\,{\footnotesize III}]}\le41.2$, while the two right columns display those with high $\log L_\mathrm{[O\,{\footnotesize III}]}>41.5$. The remaining details of this figure are the same as those in Figure \ref{fig:ex_1}.
\label{fig:ex_2}}
\end{figure*}

\begin{figure*}
\includegraphics[width=\linewidth]{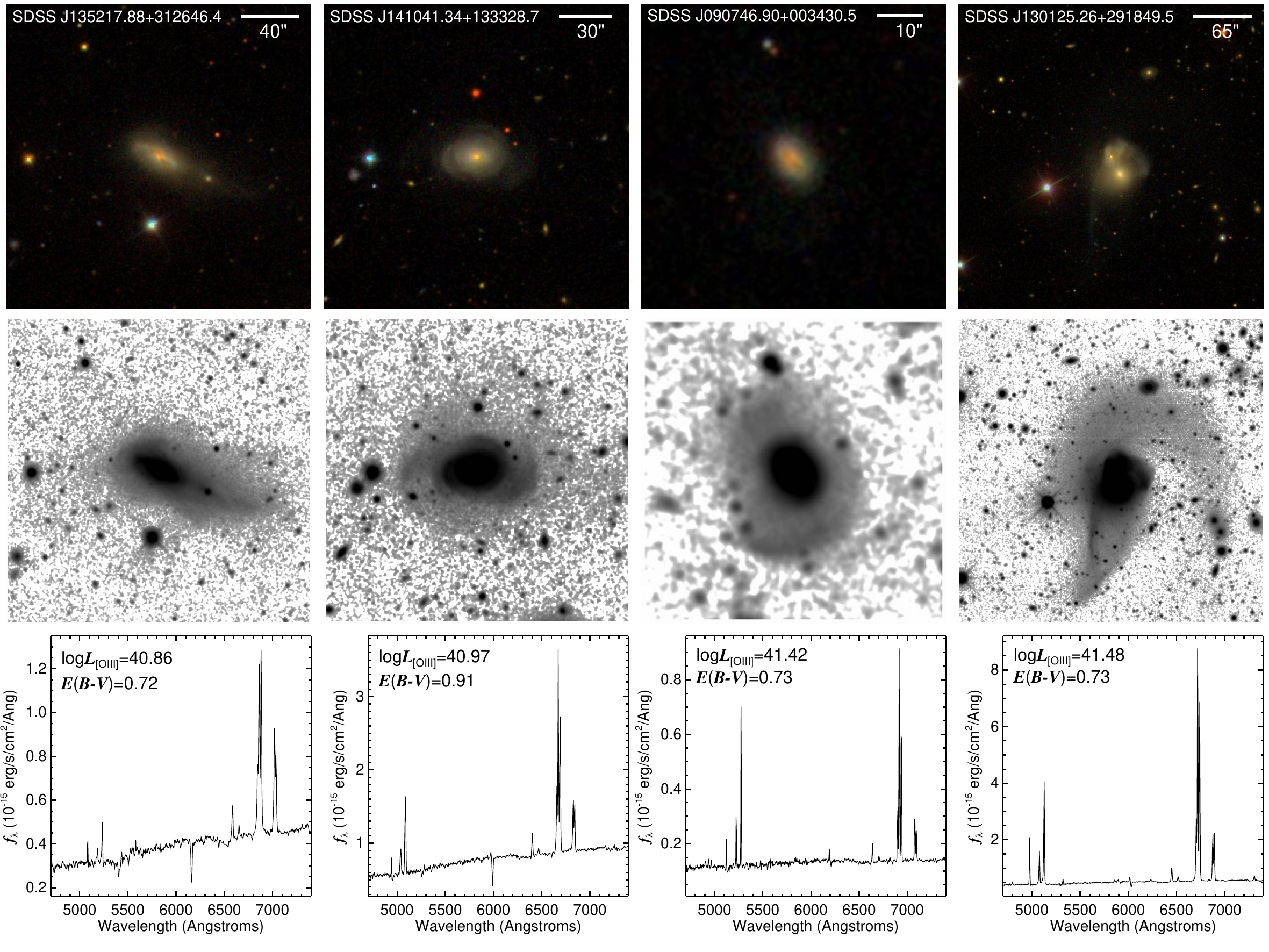}
\centering
\caption{Examples of type 2 AGN hosts with tidal features, which have high $E(B-V)>0.7$. The two left columns show AGNs with low $\log L_\mathrm{[O\,{\footnotesize III}]}<41.0$, while the two right columns display those with high $\log L_\mathrm{[O\,{\footnotesize III}]}>41.4$. The remaining details of this figure are the same as those in Figure \ref{fig:ex_1}.
\label{fig:ex_3}}
\end{figure*}

\begin{figure*}
\includegraphics[width=\linewidth]{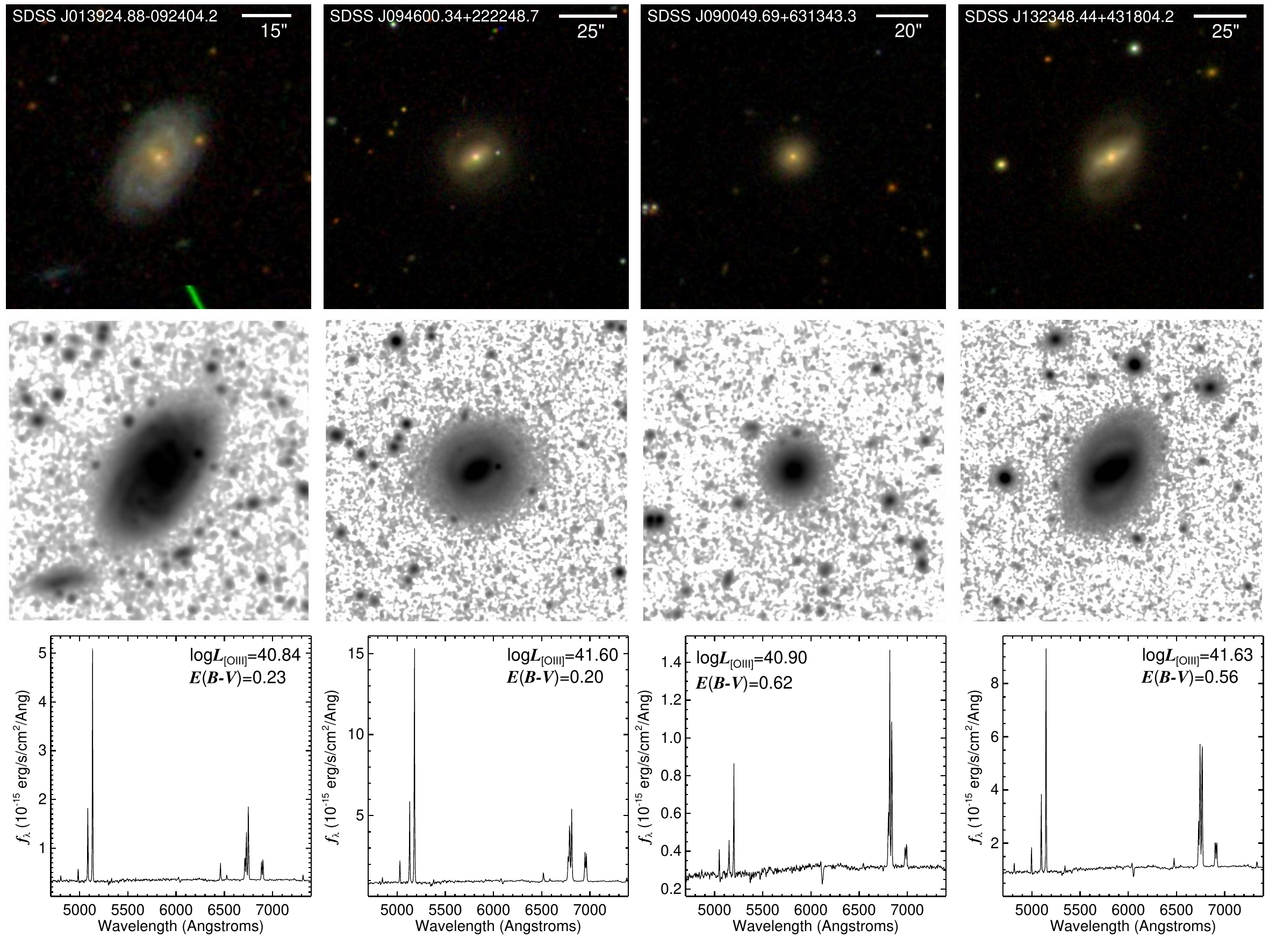}
\centering
\caption{Examples of type 2 AGN hosts that do not have tidal features. The two left columns show AGN hosts with low $E(B-V)<0.3$, while the two right columns display those with high $E(B-V)>0.5$. The first and third columns correspond to AGNs with low $\log L_\mathrm{[O\,{\footnotesize III}]}<41$, whereas the second and fourth columns correspond to AGNs with high $\log L_\mathrm{[O\,{\footnotesize III}]}>41.5$. The remaining details of this figure are the same as those in Figure \ref{fig:ex_1}.
\label{fig:ex_ntf}}
\end{figure*}

\section{Sample}\label{sec:sample}

We used type 2 AGNs at $z<0.063$, selected based on emission-line flux ratios ([O\,{\footnotesize III}] $\lambda$5007/H$\beta$ and [N\,{\footnotesize II}] $\lambda$6584/H$\alpha$) in the Baldwin, Phillips \& Terlevich (BPT) diagram \citep{Baldwin1981}. The redshift range of $z<0.063$ was adopted to balance the need for a statistically sufficient number of AGNs with the purpose of minimizing the inclusion of higher redshift AGN hosts,\footnote{As noted in Section \ref{sec:results}, the redshift range of the sample does not significantly affect our main findings.} as these tend to have smaller angular sizes and suffer from cosmological surface-brightness dimming (see Equation (6) in \citealt{YP2020}), both of which hinder the detection of tidal features.

The fluxes and other properties for emission lines, such as [O\,{\footnotesize III}] $\lambda$5007, [N\,{\footnotesize II}] $\lambda$6584, H$\beta$, and H$\alpha$ are obtained from the Max Planck Institute for Astrophysics--Johns Hopkins University (MPA--JHU) catalog,\footnote{\url{http://www.sdss.org/dr17/spectro/galaxy_mpajhu/}} which is based on Sloan Digital Sky Survey (SDSS) Data Release 8 \citep{Aihara2011}. We include only those galaxies in which the signal-to-noise ratio (S/N) of the emission lines used in the BPT diagram exceeds five for the selection of type 2 AGNs.\footnote{Our results remain essentially unchanged when applying a S/N cut of 10.} In addition, we restricted the sample to galaxies with a full width at half maximum (FWHM) of the Balmer lines less than 500 $\mathrm{km\,s}^{-1}$ \citep{Whittle1985}, which is measured simultaneously across all Balmer lines, in order to exclude galaxies with broad emission lines. We note that the SDSS pipeline classifies objects as having broad emission lines if it detects lines with $\mathrm{FWHM}\gtrsim470\, \mathrm{km\,s}^{-1}$. By cross-matching our type 2 AGN sample\footnote{The type 2 AGN sample without the FWHM cut.} with the SDSS-based type 1 AGN catalog of \citet{Oh2015}, we find that roughly half of the objects with Balmer-line $\mathrm{FWHM}>500\,\mathrm{km\,s}^{-1}$ are likely type 1 AGNs, and this fraction further increases to $\gtrsim70\%$ for $\mathrm{FWHM}\gtrsim600$-$700\,\mathrm{km\,s}^{-1}$. By contrast, the fraction remains below $7\%$ for objects with $\mathrm{FWHM}<400\,\mathrm{km\,s}^{-1}$. Therefore, adopting $\mathrm{FWHM}<500\,\mathrm{km\,s}^{-1}$ provides a conservative threshold that effectively reduces type 1 contamination without excessively diminishing the size of the type 2 AGN sample.

Figure \ref{fig:bpt} displays the distribution of galaxies at $z<0.063$ in the BPT diagram. Galaxies above the solid line in the figure, which is defined by Equation (5) in \citet{Kewley2001}, were selected as AGNs. Since this line represents the theoretical upper limit for pure starburst models, galaxies lying above it exhibit emission-line ratios that cannot be reproduced by any combination of parameters within a star-forming model, indicating that their line fluxes are substantially dominated by AGN contribution. The number of type 2 AGNs at $z<0.063$ is 5578.

We derived the dust extinction of AGNs, parameterized by the color excess $E(B-V)$, using the Balmer decrement (the flux ratio H$\alpha/$H$\beta$ \footnote{The two emission-line fluxes are corrected for Galactic extinction using the dust-reddening maps of \citet{Schlegel1998}.}). Defining the observed Balmer decrement as $D_\mathrm{obs}\equiv(\mathrm{H}\alpha/\mathrm{H}\beta)_\mathrm{obs}$ and assuming an intrinsic Balmer decrement $D_0\equiv(\mathrm{H}\alpha/\mathrm{H}\beta)_0$, $E(B-V)$ is given by
\begin{equation}
E(B-V) = \frac{2.5}{k(\mathrm{H}\beta) - k(\mathrm{H}\alpha)}
\log_{10}\left(\frac{D_{\mathrm{obs}}}{D_0}\right),
\label{eq:ebv}
\end{equation}
where $k(\lambda) \equiv A_\lambda/E(B-V)$ is adopted from the reddening law of \citet{Fitzpatrick1999}, assuming a total-to-selective extinction ratio of $R_V=3.1$. Here, $k(\mathrm{H}\beta)$ and $k(\mathrm{H}\alpha)$ denote the values of $k(\lambda)$ evaluated at the wavelengths of H$\beta$ and H$\alpha$, respectively. We assumed an intrinsic Balmer decrement of 2.86, consistently with previous studies of AGNs \citep{Kauffmann2003,Weston2017,Hernandez2023}. 

The luminosity of the [O\,{\footnotesize III}] $\lambda$5007 emission line ($L_\mathrm{[O\,{\footnotesize III}]}$) was employed as a proxy for AGN luminosity, as commonly adopted in previous studies \citep{Alonso2007,Ellison2013,Oh2015,Weston2017,Comerford2024}. In typical type 2 AGNs, the [O\,{\footnotesize III}] $\lambda$5007 emission line is one of the strongest emission lines and is less contaminated by star formation than other optical emission lines. For example, \citet{Heckman2004} suggested that the AGN contribution to $L_\mathrm{[O\,{\footnotesize III}]}$ exceeds $90\%$ in AGN-dominated emission-line galaxies selected from the BPT diagram, as done in this study, indicating that it is a reliable tracer of AGN power. 

In this study, $L_\mathrm{[O\,{\footnotesize III}]}$ was corrected for internal extinction derived from the Balmer decrement. The extinction-corrected  $L_\mathrm{[O\,{\footnotesize III}]}$ was computed as
\begin{equation}
L_\mathrm{[O\,{\footnotesize III}]} = 4\pi D_L^2 \, F_\mathrm{[O\,{\footnotesize III}]} \, 10^{0.4\,k([\mathrm{O\,{\footnotesize III}]}\,\lambda 5007)\,E(B-V)},
\label{eq:oiii}
\end{equation}
where $D_L$ is the luminosity distance at the source redshift calculated using the adopted cosmology, and $F_\mathrm{[O\,{\footnotesize III}]}$ is the observed [O\,{\footnotesize III}] flux corrected only for Galactic extinction using the dust-reddening maps of \citet{Schlegel1998}. The term $k([\mathrm{O\,{\footnotesize III}]}\,\lambda 5007)$ denotes the value of the \citet{Fitzpatrick1999} reddening curve evaluated at the wavelength of the [O\,{\footnotesize III}] $\lambda$5007 emission line, and $E(B-V)$ is derived from the Balmer decrement as given in Equation \ref{eq:ebv}.

We estimate the AGN bolometric luminosity ($L_\mathrm{bol}$) from $L_\mathrm{[O\,{\footnotesize III}]}$ using the conversion $L_\mathrm{bol}\approx600\,L_\mathrm{[O\,{\footnotesize III}]}$, which was proposed by \citet{Kauffmann2009} for extinction-corrected [O\,{\footnotesize III}] $\lambda$5007 luminosity.\footnote{Similarly, a factor of 700 was used for the bolometric correction in \citet{LaMassa2009}.} Throughout this paper, $L_\mathrm{[O\,{\footnotesize III}]}$ is given in units of erg s$^{-1}$.

In this study, we used type 2 AGNs with $\log L_\mathrm{[O\,{\footnotesize III}]}\ge40.8$.\footnote{This luminosity range ensures [O\,{\footnotesize III}] emission lines with $\mathrm{S/N}\gtrsim20$.} This yielded a sample of 787 AGNs, 39 of which were excluded due to poor image quality, primarily caused by proximity to bright sources (Section \ref{sec:tidal}). The final sample used in this study therefore consists of 748 AGNs. The black circles in Figure \ref{fig:bpt} indicate our type 2 AGNs in the BPT diagram. As indicated by different symbols and colors in Figure \ref{fig:bpt}, the typical flux ratios of AGNs across different $E(B-V)$ and $L_\mathrm{[O\,{\footnotesize III}]}$ bins do not vary significantly, particularly for [N\,{\footnotesize II}] $\lambda$6584/ H$\alpha$. This demonstrates that our samples in different bins do not occupy substantially different regions of the BPT diagram.

The requirement that AGNs in our sample have $\log L_\mathrm{[O\,{\footnotesize III}]}\ge40.8$ and S/N$>5$ in all the emission lines used in the BPT diagram effectively excludes low-ionization nuclear emission-line regions (LINERs), which may not be associated with active SMBHs. Using the LINER classification criterion from \citet{Schawinski2007}, we find that only $9\%$ of our sample can be classified as LINERs. These are included in our final sample, but excluding them does not essentially change our results.

We compare $E(B-V)$ derived from the Balmer decrement with that obtained from stellar continuum fitting, which is entirely independent of the Balmer line-based measurement, in order to gain insight into the nature of extinctions in our type 2 AGN sample. The stellar continuum fitting results are taken from the Firefly value-added catalog \citep{Comparat2017}, which uses stellar population models of \citet{Maraston2011} and a full spectral fitting code based on chi-squared minimization. Among the results from various stellar libraries and initial mass functions (IMFs) that the catalog provides, we selected those based on STELIB stellar library \citep{LeBorgne2003} with the \citet{Chabrier2003} IMF. The use of other options yields similar results. We find that 662 AGNs in our sample match the catalog.  

Figure \ref{fig:ebv} shows the comparison between $E(B-V)$ derived from the Balmer decrement and that obtained from stellar continuum fitting. The two parameters exhibit a notable correlation, with a Pearson correlation coefficient of 0.67 and a mean of the absolute deviation of 0.1 from the best-fit relation. This correlation suggests that the dust extinction of AGN emission lines is more likely associated with dust on the scale of the host galaxy, rather than with a local geometric effect from dust in the immediate vicinity of the SMBH. However, we find that $E(B-V)$ derived from the Balmer decrement is, on average, 0.35 higher than that obtained from stellar continuum fitting. This gap is reduced to only 0.28 even when adopting a higher intrinsic Balmer decrement of 3.1, as used for AGNs in several previous studies \citep{Halpern1983,Jaffarian2020,Macconi2020}, instead of 2.86. This implies that the line-of-sight extinction toward the narrow-line region near the SMBH, embedded in the galactic nucleus, is higher than the stellar light-weighted extinction of the central region covered by the fiber aperture.

We note that $E(B-V)$ derived from the Balmer decrement has several advantages for studies of type 2 AGNs compared to $E(B-V)$ inferred from stellar continuum fitting. First, the intrinsic Balmer-line ratio is set by atomic physics, and $E(B-V)$ is derived from a simple two-line ratio, making the extinction estimate far less model dependent. By contrast, stellar population modeling is susceptible to age-metallicity-dust degeneracies and depends sensitively on the assumed attenuation curve and star formation history. Second, $E(B-V)$ from the Balmer decrement directly probes the dust along the line of sight to the narrow-line region of the AGN, whereas $E(B-V)$ from stellar continuum fitting traces a luminosity-weighted average over the stellar populations within the fiber aperture, which can have a relatively weak connection to the AGN.

In the first rows of Figures \ref{fig:ex_1}--\ref{fig:ex_ntf}, we present color images of example type 2 AGN hosts, while in the third rows, we show the optical spectrum of each AGN covering the observed wavelength range of 4700--7400\AA, where the H$\beta$, [O\,{\footnotesize III}], and H$\alpha$ emission lines are visible. Table \ref{table} presents the properties of type 2 AGNs in our sample, including information on the presence of tidal features in their host galaxies.
\\

\begin{table*}[h!]
\caption{Properties of type 2 AGNs.
\label{table}}
\centering
\begin{tabular}{lccccccc}
\hline\hline
\tiny{SDSS ObjID} & \tiny{R.A.}\tablefootmark{a} & \tiny{decl}.\tablefootmark{a} & \tiny{Redshift} & \tiny{$E(B-V)$}\tablefootmark{b} & \tiny{$\log L_\mathrm{[O\,{\footnotesize III}]}$}\tablefootmark{c} & \tiny{$\log M_\mathrm{star}$}\tablefootmark{d}  & \tiny{Tidal feature}\tablefootmark{e}\\
\hline
\tiny{1237661812812349452} & \tiny{186.44468} & \tiny{ 12.66188} & \tiny{0.00862} & \tiny{0.55} & \tiny{41.57} & \tiny{10.7} & \tiny{Y}\\
\tiny{1237660024523849823} & \tiny{ 45.95464} & \tiny{ -1.10373} & \tiny{0.01363} & \tiny{0.62} & \tiny{40.85} & \tiny{10.9} & \tiny{Y}\\
\tiny{1237667142857916600} & \tiny{119.91716} & \tiny{ 15.38682} & \tiny{0.01552} & \tiny{0.62} & \tiny{41.16} & \tiny{10.5} & \tiny{N}\\
\tiny{1237662530065465417} & \tiny{212.67229} & \tiny{ 13.55800} & \tiny{0.01622} & \tiny{0.91} & \tiny{40.97} & \tiny{10.3} & \tiny{Y}\\
\tiny{1237649920576258107} & \tiny{ 14.91714} & \tiny{ 15.33098} & \tiny{0.01829} & \tiny{0.39} & \tiny{40.92} & \tiny{10.6} & \tiny{N}\\
\tiny{1237664836462182621} & \tiny{124.90783} & \tiny{ 21.11430} & \tiny{0.01848} & \tiny{0.49} & \tiny{41.01} & \tiny{10.9} & \tiny{N}\\
\tiny{1237652949068284053} & \tiny{ 13.37466} & \tiny{ -8.76779} & \tiny{0.01898} & \tiny{0.48} & \tiny{41.38} & \tiny{10.6} & \tiny{N}\\
\tiny{1237661068722372615} & \tiny{175.57032} & \tiny{ 14.06660} & \tiny{0.02074} & \tiny{0.21} & \tiny{41.13} & \tiny{10.2} & \tiny{N}\\
\tiny{1237667323784593449} & \tiny{161.24281} & \tiny{ 26.09143} & \tiny{0.02094} & \tiny{0.37} & \tiny{41.03} & \tiny{ 9.8} & \tiny{N}\\
\tiny{1237665532252323890} & \tiny{222.65772} & \tiny{ 22.73434} & \tiny{0.02095} & \tiny{0.18} & \tiny{41.51} & \tiny{10.1} & \tiny{N}\\
\tiny{1237668624630611992} & \tiny{203.02007} & \tiny{ 17.04897} & \tiny{0.02165} & \tiny{0.28} & \tiny{40.86} & \tiny{10.7} & \tiny{N}\\
\tiny{1237659348026589196} & \tiny{200.05655} & \tiny{  7.90781} & \tiny{0.02169} & \tiny{0.40} & \tiny{40.82} & \tiny{ 9.9} & \tiny{N}\\
\tiny{1237668297138045048} & \tiny{203.48241} & \tiny{ 18.30590} & \tiny{0.02247} & \tiny{0.50} & \tiny{40.85} & \tiny{10.4} & \tiny{N}\\
\tiny{1237663277925990500} & \tiny{356.00858} & \tiny{  0.51661} & \tiny{0.02251} & \tiny{0.93} & \tiny{41.26} & \tiny{10.6} & \tiny{N}\\
\tiny{1237667915955634219} & \tiny{181.12362} & \tiny{ 20.31630} & \tiny{0.02262} & \tiny{0.37} & \tiny{41.58} & \tiny{10.7} & \tiny{N}\\
\hline
\end{tabular}
\tablefoot{The full table is available at the CDS.\\
\tablefoottext{a}{R.A. and decl. are in degrees.}
\tablefoottext{b}{The color excess $E(B-V)$ derived from the Balmer decrement.}
\tablefoottext{c}{The logarithmic luminosity of the [O\,{\footnotesize III}] $\lambda$5007 emission line. Here, $L_\mathrm{[O\,{\footnotesize III}]}$ is corrected for internal extinction derived from the Balmer decrement. $L_\mathrm{[O\,{\footnotesize III}]}$ is expressed in units of erg s$^{-1}$.}
\tablefoottext{d}{The logarithmic stellar mass obtained from the MPA--JHU catalog. $M_\mathrm{star}$ is expressed in solar masses ($M_{\odot}$).}
\tablefoottext{e}{The presence of tidal features in the host galaxy: Y = present, N=absent.}
}
\end{table*}

\section{Detection of tidal features}\label{sec:tidal}

The method and deep images used for detecting of tidal features are identical to those in \citet{Yoon2025}. To detect tidal features, we used images from Data Release 10 of the DESI Legacy Imaging Survey \citep{Dey2019}, which combines three wide-area surveys: the Dark Energy Camera Legacy Survey, the Beijing--Arizona Sky Survey, and the Mayall $z$-band Legacy Survey. Collectively, these surveys cover $\sim14,000$ square degrees of the sky. The DESI Legacy Imaging Survey achieves a median surface-brightness depth of $\sim27$ mag arcsec$^{-2}$ in the $g$ and $r$ bands, based on the $1\sigma$ noise level measured over a $1\arcsec\times1\arcsec$ area. This level of surface-brightness depth is comparable to that achieved in the deep, coadded $r$-band images of the SDSS Stripe 82 region \citep{YL2020,Yoon2022,Yoon2023}, which have been widely used to detect low-surface-brightness tidal features around galaxies (e.g., \citealt{Kaviraj2010,Schawinski2010,Hong2015}).

The identification of tidal features is performed through visual inspection of individual $g$- and $r$- band images, in combination with composite color images generated from the $g$, $r$, and $z$ bands. To better detect low-surface-brightness tidal features, we applied two display adjustments during the visual inspection process. First, intensity scaling was used. The images are displayed with a logarithmic scale, and the minimum and maximum display levels are adjusted to optimize the visibility of faint, diffuse structure. Second, Gaussian smoothing was applied using kernels with widths in the range of $\sigma\approx1$-$3$ pixels to suppress small-scale noise and improve the detectability of extended, low-contrast tidal features. These procedures are applied interactively and manually during the visual inspection, with the intensity scaling and smoothing parameters adjusted for each image individually, because the brightness and spatial extent of tidal features vary from object to object. Visual inspection is a widely used and reliable method for detecting tidal features and other merger-related disturbances (e.g., \citealt{Darg2010,Cisternas2011,Treister2012,Kocevski2015,Ellison2019}), and it is considered the standard of comparison for evaluating quantitative techniques used to identify merger features (e.g., \citealt{Conselice2003,Lotz2004}).

Tidal features manifest in various forms, commonly classified as streams, tails, and shells \citep{Duc2015,Mancillas2019,Bilek2020,Bilek2023,Sola2022}. Tidal shells appear as arc-shaped structures with sharp edges. Depending on the system, these arcs can exhibit a coherent alignment or be distributed irregularly around the host galaxy. Shells at larger distances from the host galaxy typically appear more diffuse. Displaying narrow and elongated structures, tidal streams are commonly associated with minor merger events. In some cases, these streams extend from or are directly connected to nearby satellite galaxies, indicating ongoing or recent merger processes. Tidal tails are wide, elongated stellar structures that emerge prominently from host galaxies, often forming as a result of major mergers. While they resemble tidal streams in morphology, tidal tails are distinguished by their larger spatial extent, which can be comparable to the size of the host galaxy itself. Nonetheless, in certain systems, the distinction between tails and streams may be ambiguous due to their shared properties. A number of galaxies exhibit a mix of different types of tidal features. In this study, tidal tails, streams, and shells observed around galaxies are collectively classified as tidal features.\footnote{For example, the AGN host in the first column of Figure \ref{fig:ex_3} shows a tidal stream. The AGN host in the second column of Figure \ref{fig:ex_1} exhibits tidal tails, while the one in the third column of Figure \ref{fig:ex_3} displays a tidal shell.}

We identify tidal features in 161 out of 748 type 2 AGN hosts, corresponding to $21.5\%$ of the sample. Examples of $r$-band deep images from the DESI Legacy Imaging Survey for type 2 AGN hosts with tidal features are presented in the second rows of Figures \ref{fig:ex_1}--\ref{fig:ex_3}, while those for type 2 AGN hosts without tidal features are shown in the second row of Figure \ref{fig:ex_ntf}.

The reliability of our method for identifying tidal features in AGN hosts is assessed by comparing our primary classifications (performed by Y.Y.) with an independent classification conducted by J.C. on a subsample of 263 type 2 AGNs with $\log L_\mathrm{[O\,{\footnotesize III}]}\gtrsim41.4$ or $E(B-V)\gtrsim0.6$, a subset that shows a high fraction of AGN hosts with tidal features. The comparison reveals that $94.2\%$ (81/86) of AGN hosts classified as exhibiting tidal features by Y.Y. are likewise identified as such by J.C., whereas $93.2\%$ (165/177) of AGN hosts without tidal features in the Y.Y. classification are identified as having no tidal features by J.C. This indicates that the detection of tidal features is highly consistent between different identifiers, despite its reliance on visual inspection. 

 A similarly high level of agreement---around $90\%$---for tidal feature identification is also found in our previous studies on samples of type 1 AGNs and early-type galaxies \citep{YL2020,Yoon2022,Yoon2024b,Yoon2025}. Given the accuracy rate of $\gtrsim90\%$, an additional uncertainty of about 0.01 to 0.08 may be introduced into the tidal feature fractions reported in Section \ref{sec:results}, depending on the number of galaxies in each bin. Nonetheless, this level of uncertainty remains smaller than the standard errors adopted in this study.
\\

\begin{figure}
\includegraphics[width=\linewidth]{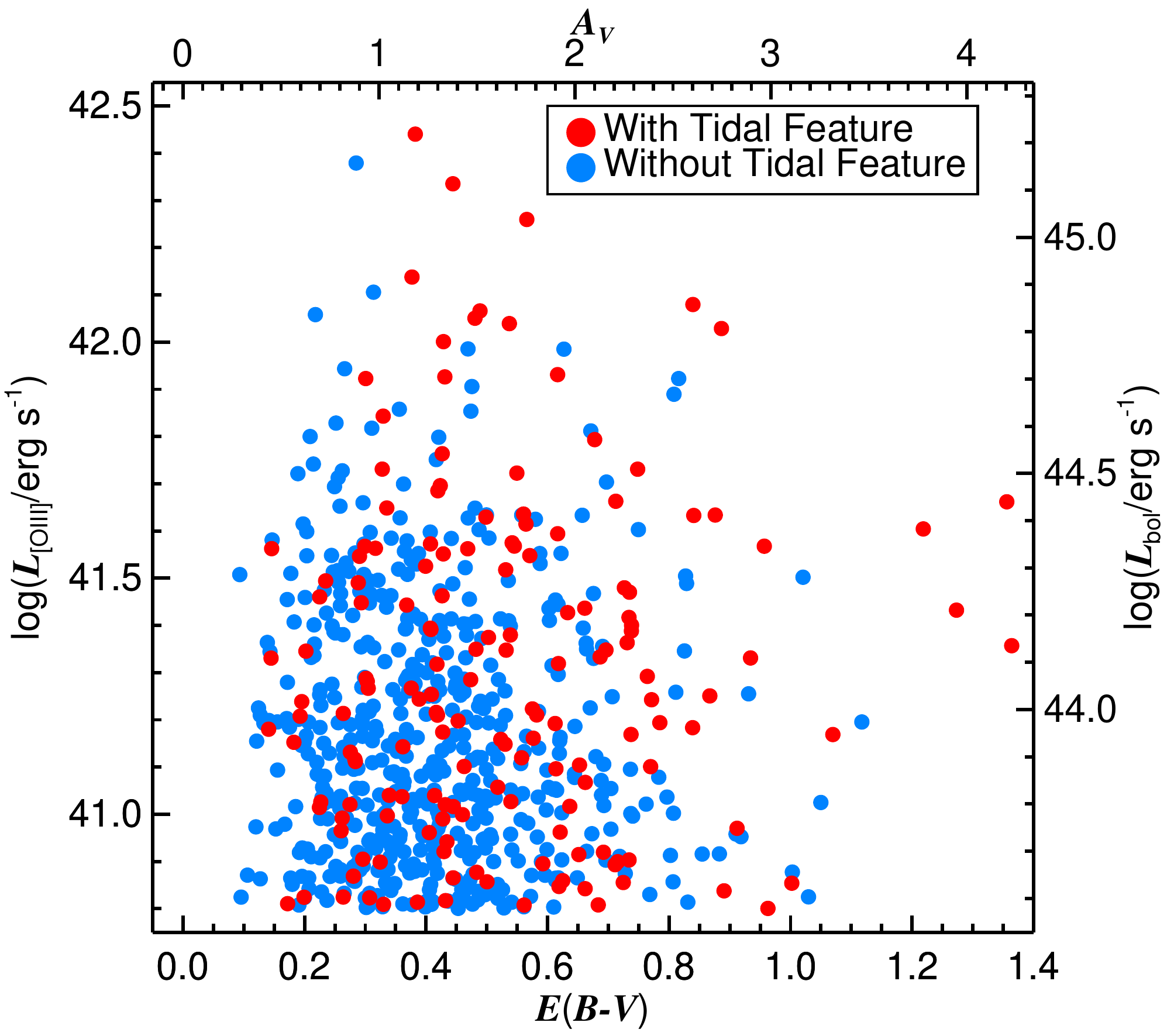}
\centering
\caption{Distribution of type 2 AGNs in the plane of $E(B-V)$ (or $A_V$) versus $\log L_\mathrm{[O\,{\footnotesize III}]}$ (or $\log L_\mathrm{bol}$). AGNs are divided into two categories based on the presence or absence of tidal features.
\label{fig:2ddist}}
\end{figure} 

\begin{figure}
\includegraphics[width=\linewidth]{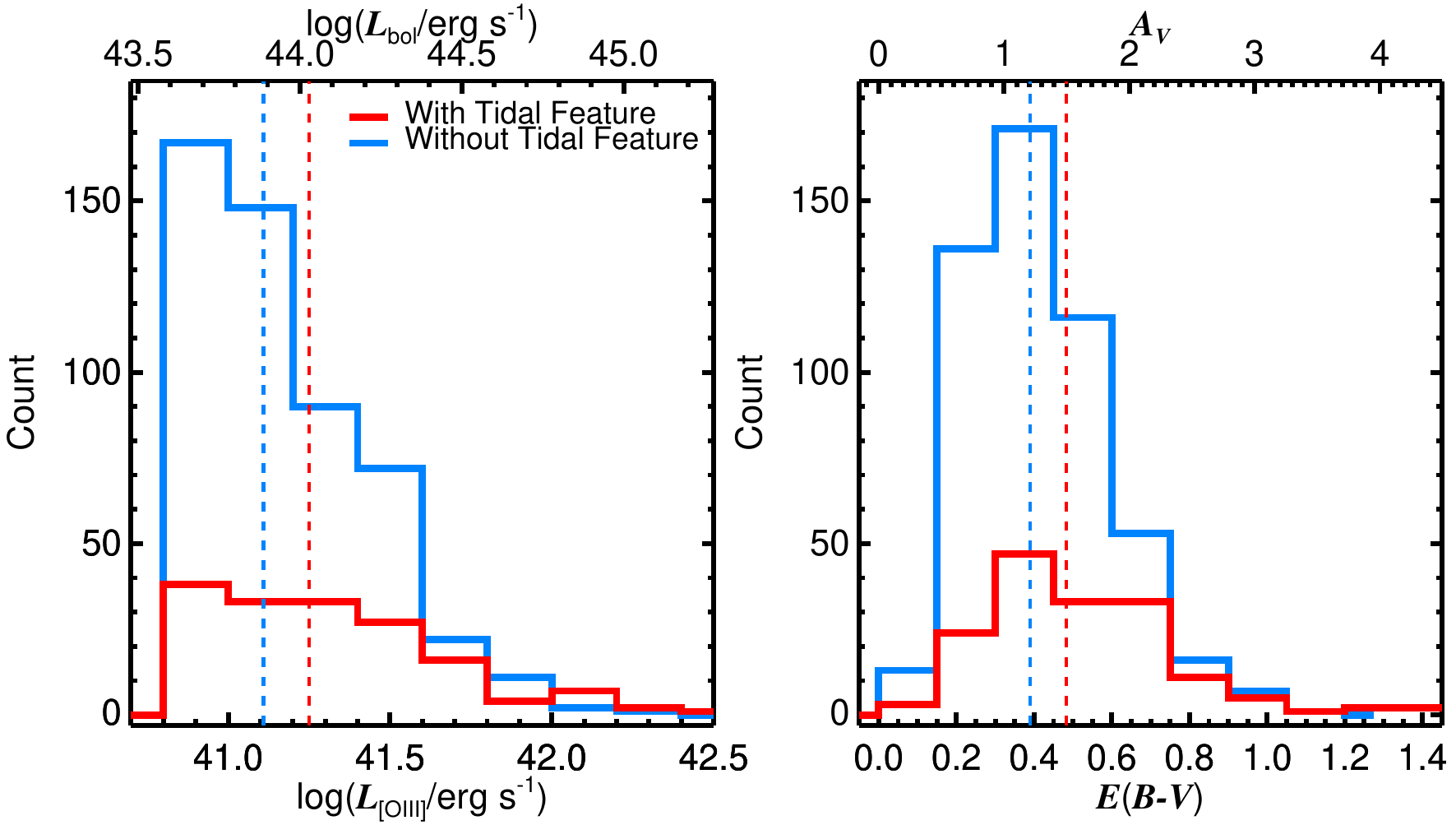}
\centering
\caption{Distributions of $\log L_\mathrm{[O\,{\footnotesize III}]}$ ($\log L_\mathrm{bol}$) and $E(B-V)$ ($A_V$) for AGN hosts with and without tidal features. The vertical dashed lines mark the median value of each parameter for the two AGN categories.
\label{fig:dist}}
\end{figure} 

\begin{figure*}
\includegraphics[scale=0.23]{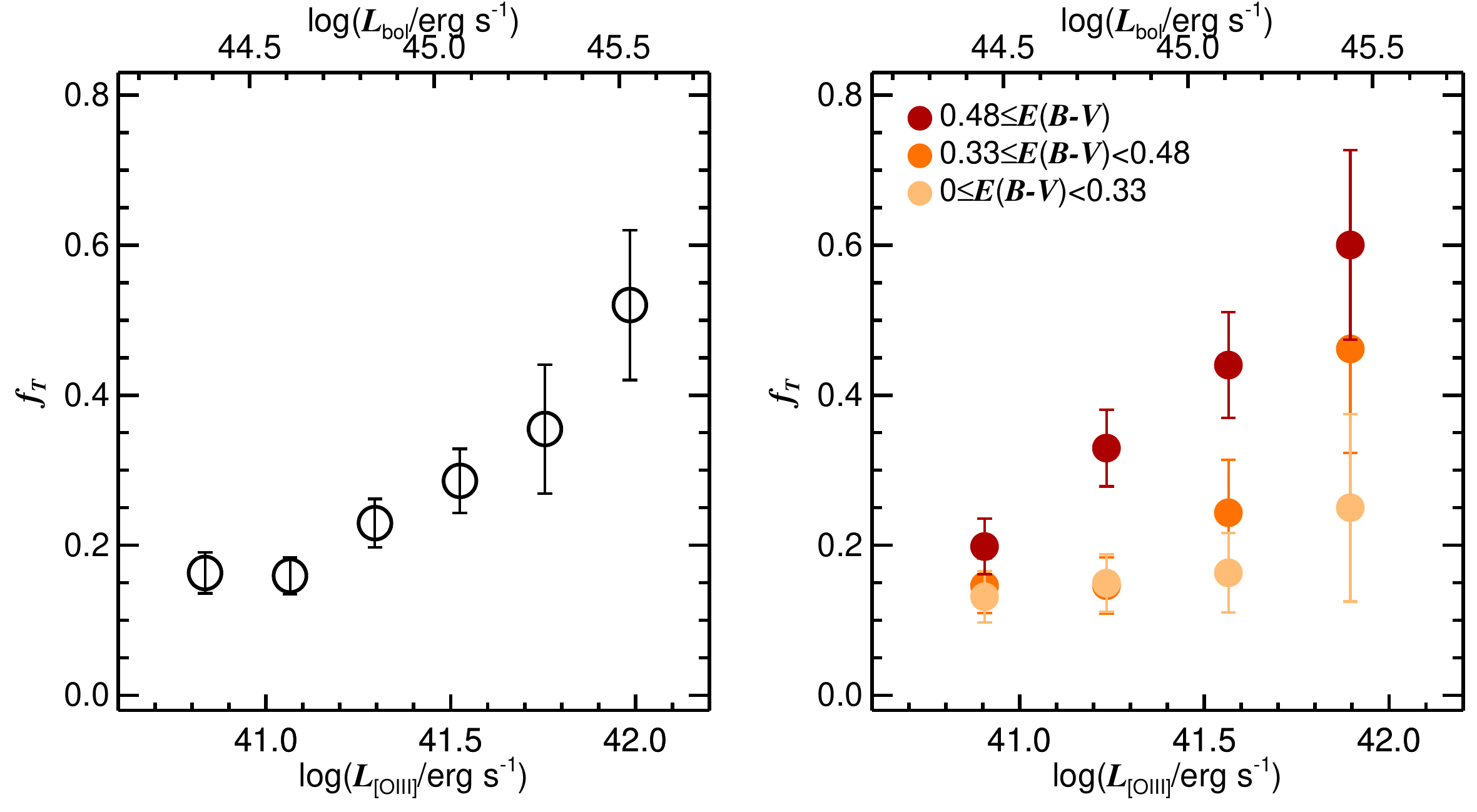}\includegraphics[scale=0.23]{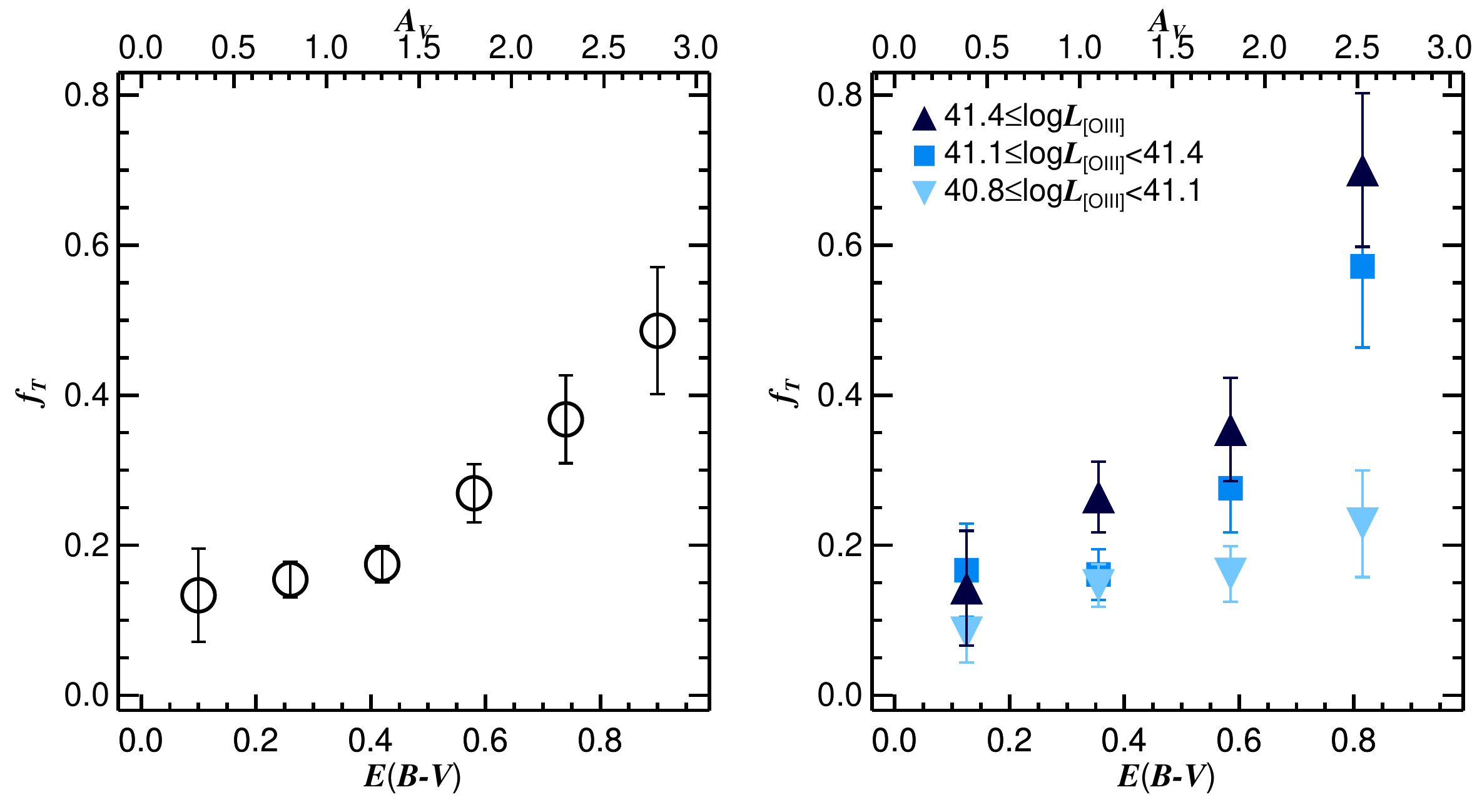}
\centering
\caption{First and second panels: Fraction of type 2 AGN hosts with tidal features ($f_T$) as a function of $\log L_\mathrm{[O\,{\footnotesize III}]}$. The first panel exhibits the result for the full AGN sample, while the second panel presents the results divided into three $E(B-V)$ bins. A bin size of 0.23 dex is used in the left panel, whereas a bin size of 0.33 dex is applied in the right panel. Third and fourth panels: $f_T$ as a function of $E(B-V)$. The third panel displays the result for the full AGN sample, while the fourth panel shows the results divided into three $\log L_\mathrm{[O\,{\footnotesize III}]}$ bins. A bin size of 0.16 is used in the left panel, while a bin size of 0.23 is applied in the right panel. In all the panels, the error bars represent the standard error of the proportion, and the highest bins are overflow bins that include all values above the upper bound of the defined range, thereby capturing the extremes of the AGN population.
\label{fig:frac}}
\end{figure*}

\begin{figure}
\includegraphics[width=\linewidth]{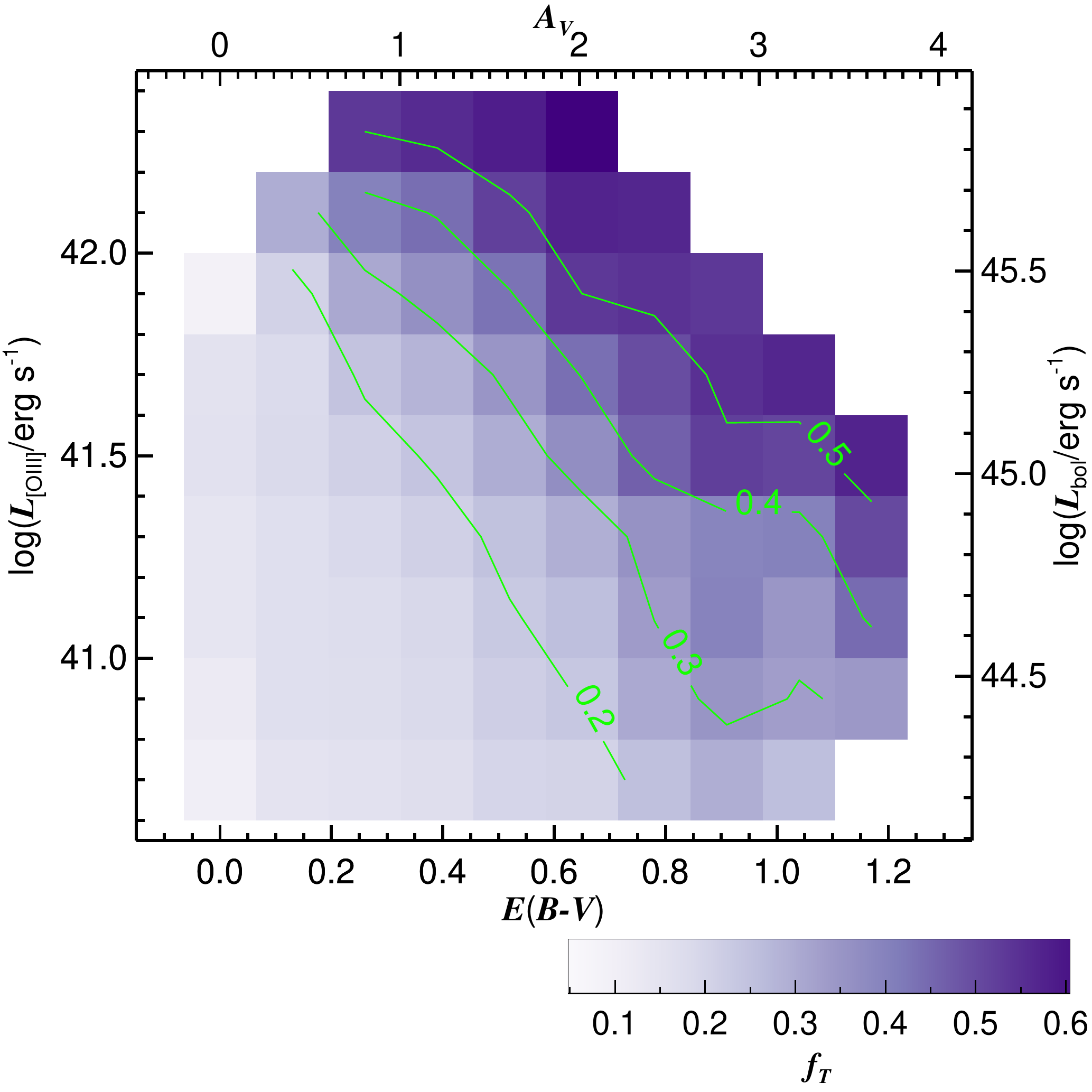}
\centering
\caption{Fraction of type 2 AGN hosts with tidal features ($f_T$) is shown by colors and contours in the $E(B-V)$ versus $\log L_\mathrm{[O\,{\footnotesize III}]}$ plane. The color bar indicates the color-coded values of $f_T$, and the numbers in the middle of the contour lines represent the corresponding $f_T$ values. To generate the color map and contours, we applied a grid with block sizes of 0.13 along the $E(B-V)$ axis and 0.20 dex along the $\log L_\mathrm{[O\,{\footnotesize III}]}$ axis. At each grid point, $f_T$ is calculated using a rectangular bin with side lengths of 0.52 and 0.80 dex along the $E(B-V)$ and $\log L_\mathrm{[O\,{\footnotesize III}]}$ axes, respectively. Using bin sizes larger than the grid-block size to compute the parameter smooths the color map and contours, thereby enhancing the visibility of large-scale trends. Only the colored bins for which more than ten AGNs were used in the calculation of $f_T$ are displayed.
\label{fig:2df}}
\end{figure}

\section{Results}\label{sec:results}

Figure \ref{fig:2ddist} presents the distribution of type 2 AGNs in the $E(B-V)$ (or $A_V$) versus $\log L_\mathrm{[O\,{\footnotesize III}]}$ (or $\log L_\mathrm{bol}$) plane, where AGNs are separated based on whether they exhibit tidal features. Figure \ref{fig:dist} displays histograms of $\log L_\mathrm{[O\,{\footnotesize III}]}$ and $E(B-V)$ for AGN hosts with and without tidal features. 

These figures show that type 2 AGN hosts with tidal features have a median $\log L_\mathrm{[O\,{\footnotesize III}]}$  0.14 dex higher than that of hosts without tidal features, which is consistent with the result found for type 1 AGN hosts in \citet{Yoon2025}. The figures also indicate that type 2 AGN hosts with tidal features have a median $E(B-V)$ value 0.09 higher than the counterparts without such features. To assess whether the $\log L_\mathrm{[O\,{\footnotesize III}]}$ distributions of AGN hosts with and without tidal features originate from the same parent population, we performed a Kolmogorov--Smirnov (KS) test. The resulting $p$ value of $2.4\times10^{-4}$ strongly rejects the null hypothesis. A similar analysis of the $E(B-V)$ distributions using a KS test yielded a highly significant $p$ value of $1.1\times10^{-6}$. Taken together, these tests provide strong evidence that type 2 AGN hosts with tidal features differ from those without, in terms of both $L_\mathrm{[O\,{\footnotesize III}]}$ and $E(B-V)$ distributions. We note that the median $A_V$ for our sample is 1.3, which is consistent with the median value of 1.1 reported for AGN hosts studied in \citet{Zhuang2020}.

The top panels of Figure \ref{fig:frac} display the fraction of type 2 AGN hosts with tidal features ($f_T$) as a function of $\log L_\mathrm{[O\,{\footnotesize III}]}$. Here, we define $f_T$ as the ratio of the number of AGN hosts exhibiting tidal features ($N_T$) to the total number of AGN hosts in a bin ($N_\mathrm{AGN}$), such that $f_T=N_T/N_\mathrm{AGN}$. The top left panel of the figure shows a clear rising trend in $f_T$ with increasing $L_\mathrm{[O\,{\footnotesize III}]}$ for the full AGN samples. For the lowest luminosity AGNs with $\log L_\mathrm{[O\,{\footnotesize III}]}<41.0$, $f_T$ is $0.16\pm0.03$, whereas for the most luminous AGNs with $\log L_\mathrm{[O\,{\footnotesize III}]}>41.9$, $f_T$ rises to $0.52\pm0.10$. A nearly identical trend is observed for type 1 AGNs in \citet{Yoon2025}, which supports the AGN unification model. 

We divided our AGN sample into three $E(B-V)$ bins and display the trend for each bin in the top right panel of Figure \ref{fig:frac}. For highly obscured AGN hosts with $E(B-V)\ge0.48$, we find an increasing trend in $f_T$ as a function of $\log L_\mathrm{[O\,{\footnotesize III}]}$; that is, it rises from $f_T=0.20$ to $0.60$. However, this trend weakens as $E(B-V)$ decreases, such that for less-obscured AGN hosts with $E(B-V)<0.33$, the $f_T$ values remain consistent within errors ($f_T=0.13$--$0.25$) across the full range of $\log L_\mathrm{[O\,{\footnotesize III}]}$.

We assessed the statistical significance of our findings using Fisher's exact test. The null hypothesis assumes no difference in $f_T$ between AGNs in the highest and lowest luminosity bins, whereas the alternative hypothesis suggests that $f_T$ is higher for AGNs in the highest luminosity bin. From the tests, we obtain a $p$ value of $1.9\times10^{-3}$ for highly obscured AGN hosts with $E(B-V)\ge0.48$ and a $p$ value of 0.24 for less-obscured AGNs with $E(B-V)<0.33$, which demonstrates that the increasing trend in $f_T$ as a function of $L_\mathrm{[O\,{\footnotesize III}]}$ is only significant for highly obscured AGN hosts.

The bottom panels of Figure \ref{fig:frac} show the $f_T$ of type 2 AGN hosts as a function of $E(B-V)$. The bottom left panel of the figure reveals a trend whereby $f_T$ increases with $E(B-V)$ for the full AGN sample. In the lowest bin ($E(B-V)<0.18$), $f_T$ is $0.13\pm0.06$, while in the highest bin ($E(B-V)\ge0.82$), $f_T$ increases substantially to $0.49\pm0.08$.

The AGN sample is divided into three bins of $L_\mathrm{[O\,{\footnotesize III}]}$, and the corresponding trends for each luminosity bin are presented in the bottom right panel of Figure \ref{fig:frac}. For high-luminosity AGNs with $\log L_\mathrm{[O\,{\footnotesize III}]}\ge41.4$, $f_T$ exhibits a clear increasing trend with $E(B-V)$, rising from $f_T=0.14$ to $0.70$. However, this trend becomes less pronounced with decreasing $\log L_\mathrm{[O\,{\footnotesize III}]}$. For low-luminosity AGNs with $\log L_\mathrm{[O\,{\footnotesize III}]}<41.1$, the $f_T$ values remain relatively constant across the full range of $E(B-V)$, varying only between 0.08 and 0.23, which is not statistically significant given the associated uncertainties. We note that AGN hosts with the lowest obscuration ($E(B-V)<0.24$) exhibit nearly identical $f_T$ values, regardless of $\log L_\mathrm{[O\,{\footnotesize III}]}$.

To evaluate the statistical significance of our results, we applied Fisher's exact test. The null hypothesis assumes no difference in $f_T$ between AGN hosts in the highest and lowest $E(B-V)$ bins, while the alternative hypothesis is that $f_T$ is higher for AGN hosts in the highest $E(B-V)$ bin. The tests yielded a $p$ value of $3.6\times10^{-4}$ for high-luminosity AGNs with $\log L_\mathrm{[O\,{\footnotesize III}]}\ge41.4$ and a $p$ value of 0.06 for low-luminosity AGNs with $\log L_\mathrm{[O\,{\footnotesize III}]}<41.1$, indicating that the increasing trend in $f_T$ with $E(B-V)$ is highly significant only for high-luminosity AGNs.

Figure \ref{fig:2df} displays $f_T$ and its variation across the $E(B-V)$ versus $\log L_\mathrm{[O\,{\footnotesize III}]}$ plane, illustrated  using colors and contours to show the combined dependence of $f_T$ on both $E(B-V)$ and $\log L_\mathrm{[O\,{\footnotesize III}]}$. This figure clearly demonstrates our main result that $f_T$ is higher for AGNs that are simultaneously luminous and obscured. In other words, such AGNs are more closely associated with galaxy mergers.

\begin{figure*}
\includegraphics[scale=0.25]{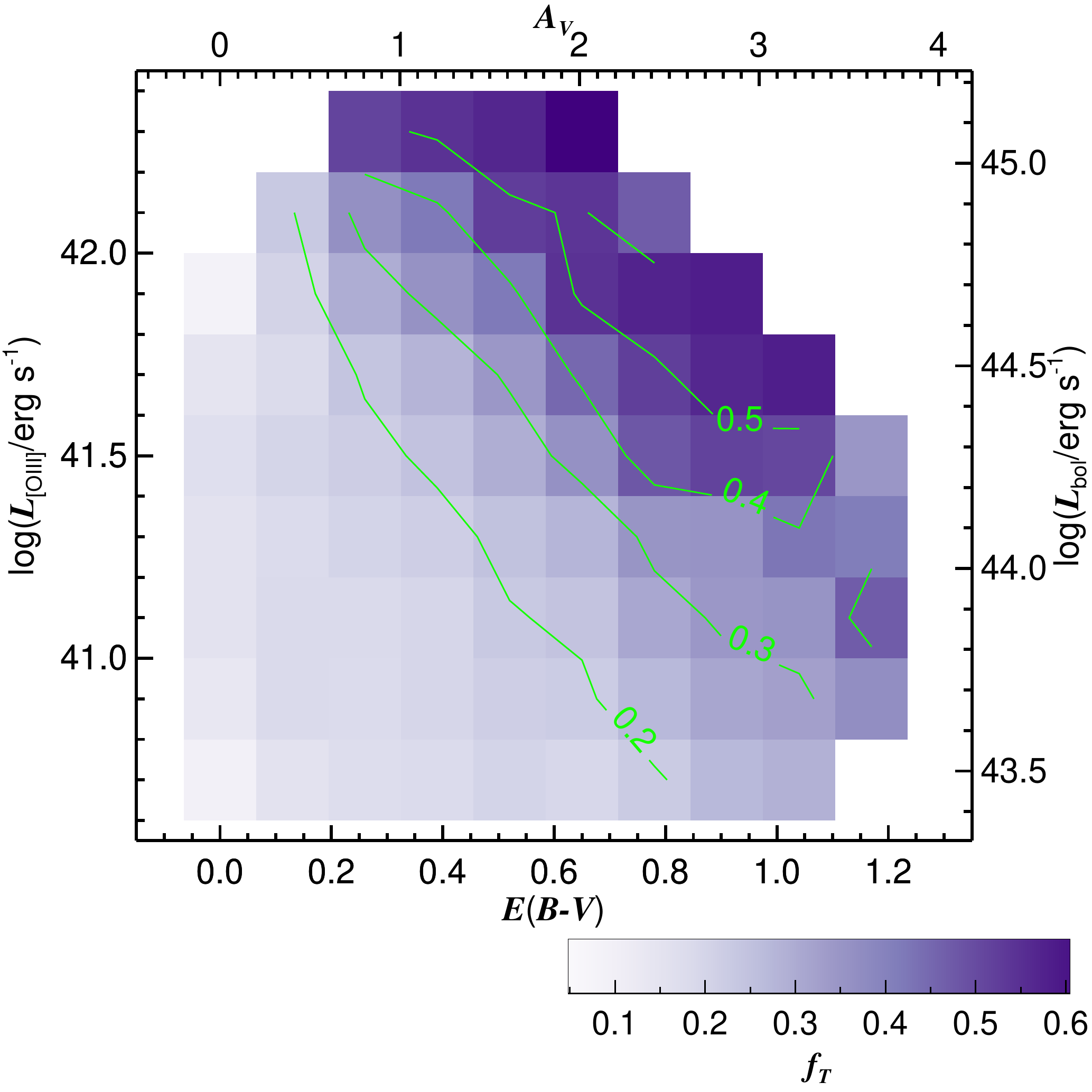}\includegraphics[scale=0.25]{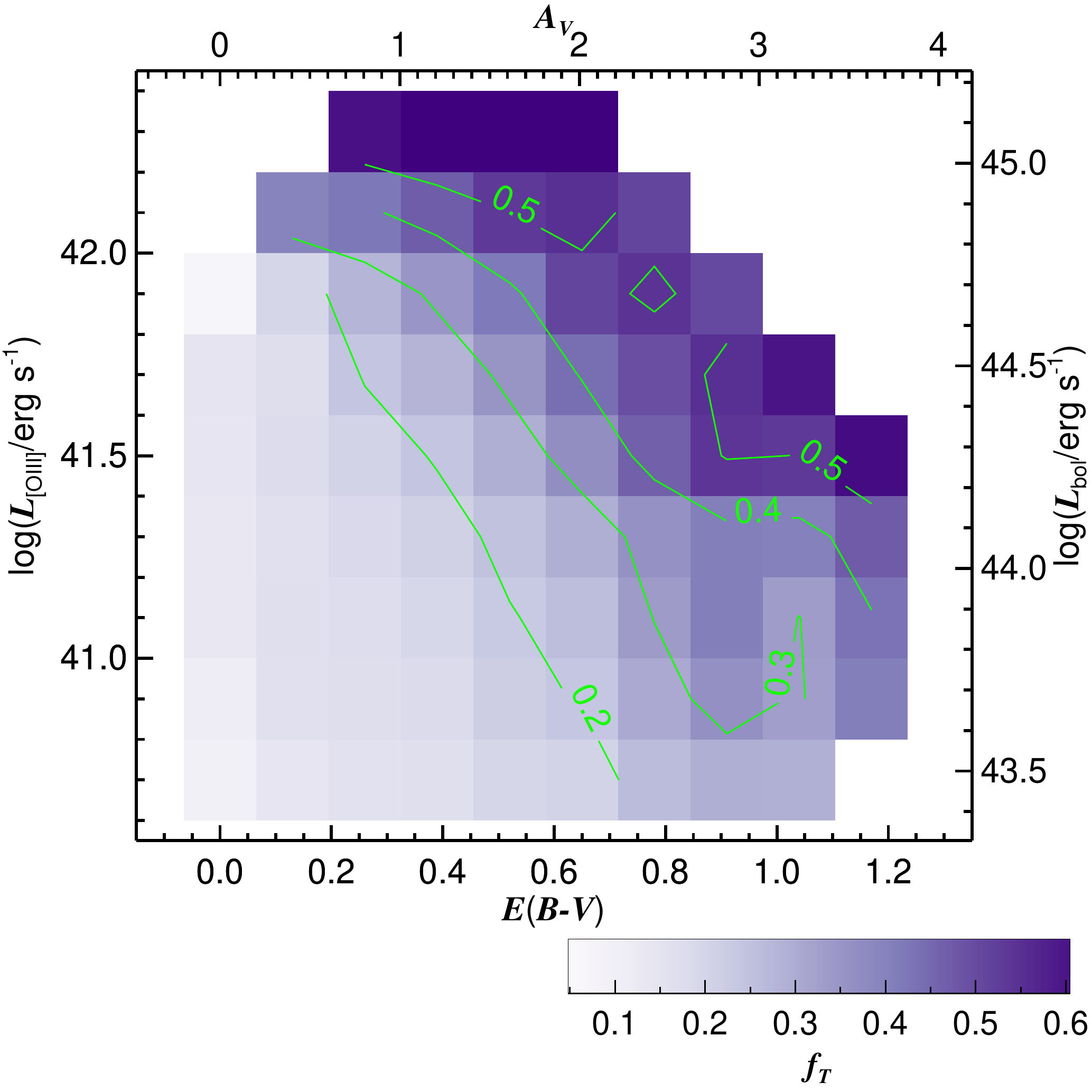}
\centering
\caption{Same as Figure \ref{fig:2df}, but with AGNs in all the $E(B-V)$ and $L_\mathrm{[O\,{\footnotesize III}]}$ bins resampled to have identical host-galaxy stellar mass distributions (the left panel) and redshift distributions (the right panel). The resampling was performed to match the stellar mass or redshift distribution of the full type 2 AGN sample, which was constructed using four bins.
\label{fig:2df_m}}
\end{figure*}

All of the results presented here are similarly reproduced when AGNs in all the $E(B-V)$ and $L_\mathrm{[O\,{\footnotesize III}]}$ bins are resampled to have identical host-galaxy stellar mass\footnote{The stellar mass information is also obtained from the MPA--JHU catalog. These stellar masses were derived from the $u$-, $g$-, $r$-, $i$-, and $z$-band galaxy photometry using a Bayesian methodology.} or redshift distributions, which is achieved through repeated sampling within each bin (see Figure \ref{fig:2df_m}). Similar tends are also observed when the AGN sample is divided into two subsamples based on either the median stellar mass of $10^{10.6}\,M_{\odot}$ or the median redshift of 0.05, although the associated uncertainties increase substantially due to limited sample sizes. This implies that neither redshift nor stellar mass is a significant factor influencing the observed trends.

\section{Discussion}\label{sec:discuss}
\subsection{Connection between AGNs and galaxy mergers}\label{sec:agnmerger}
In many previous studies \citep{Sanders1988,Cattaneo2005,Hopkins2008,Urrutia2008,Urruita2012,Banerji2012,Glikman2012,Glikman2015,Kim2015b,Kim2024a,Kim2024b,Ricci2017,KI2018}, the evolution of AGNs is interpreted within a framework initiated by gas-rich mergers that both trigger intense starbursts and fuel the central SMBHs. In the early stage of this evolutionary sequence, substantial amounts of dust and gas enshroud the nucleus, obscuring the intense radiation from the AGN.  However, as AGN feedback blows out the surrounding material, the AGN becomes unobscured. 

If this evolutionary scenario is correct and uniformly applicable to our type 2 AGN sample, the decrease in $E(B-V)$ can be assumed to represent the passage of time along the evolutionary sequence. Meanwhile, according to the evolutionary scenario, the available gas in the host galaxy is depleted over time, leading to its gradual transition into a quiescent galaxy. Consequently, a corresponding decline in AGN luminosity is naturally expected to occur alongside with the decrease in $E(B-V)$. If we adopt an additional assumption---that the timescale of AGN activity is comparable to that of tidal features, or at least the total duration of intermittent AGN episodes in the post-merger phase (despite each individual episode being short-lived) is similar to the fading timescale of tidal features---then our observational result, showing that $f_T$ decreases with decreasing $E(B-V)$ and $L_\mathrm{[O\,{\footnotesize III}]}$, supports the merger-initiated evolution model for AGNs described above. In this case, AGNs that are simultaneously obscured and luminous are more closely associated in time with (events temporally closer to) merger events.
However, the above interpretation has critical weaknesses. First, although AGN activity can last up to $\sim1$--$2$ Gyr after mergers \citep{Volonteri2015,Byrne2023} or may take more than 1 Gyr for external gas from mergers to infall into the galaxy and accrete onto the SMBH \citep{Choi2024}---a timescale comparable to the visible timescales of tidal features---the typical lifetimes of AGNs, particularly of luminous ones ($\lesssim0.1$ Gyr; \citealt{DiMatteo2005,Hopkins2005,Conroy2013,Cen2015,Schawinski2015}), are likely shorter than the fading timescales of tidal features ($\sim1$--$4$ Gyr; \citealt{Ji2014,Mancillas2019,YL2020}). Second, there is no guarantee that all type 2 AGNs in our sample uniformly follow the merger-initiated evolution model, as AGNs, especially low-luminosity ones, can be triggered through diverse mechanisms other than galaxy mergers \citep{Ohta2007,Hirschmann2012,Menci2014,Tremblay2016,Poggianti2017}.

If the typical timescale of tidal features is significantly longer than that of AGNs, then as an alternative interpretation, our result, whereby $f_T$ increases with increasing $E(B-V)$ and $L_\mathrm{[O\,{\footnotesize III}]}$, suggests that galaxy mergers preferentially trigger AGNs that are simultaneously luminous and heavily dust obscured. In other words, high-luminosity type 2 AGNs without dust obscuration and dust-obscured AGNs with low luminosities are not strongly related to merger-driven triggering. In this case, AGNs that are simultaneously obscured and luminous are more closely associated with merger events in terms of their origin than other AGNs. The two interpretations proposed here can complement each other in explaining our result rather than being mutually exclusive.\footnote{Some AGNs may be triggered by non-merger processes, yet their host galaxies may have recently undergone gas-poor mergers that produce tidal features. Such systems can be contaminants in both interpretations.} 

Previous studies have reported inconsistent results regarding the merger--AGN connection and its dependence on luminosity, as mentioned in Section \ref{sec:intro}. Our results suggest that the disagreement among different studies can originate from biases in sample selection. For example, using a sample that preferentially consists of less-obscured AGNs can result in no detection of the luminosity dependence in the merger--AGN connection, as shown in Figure \ref{fig:frac}. This implies that the selection method of the AGN sample and the ranges of AGN properties are critical factors in AGN studies.\footnote{Photometric depth can also be an notable factor in this study. We will verify our results using deeper images from future imaging surveys.}
\\

\begin{figure}
\includegraphics[width=\linewidth]{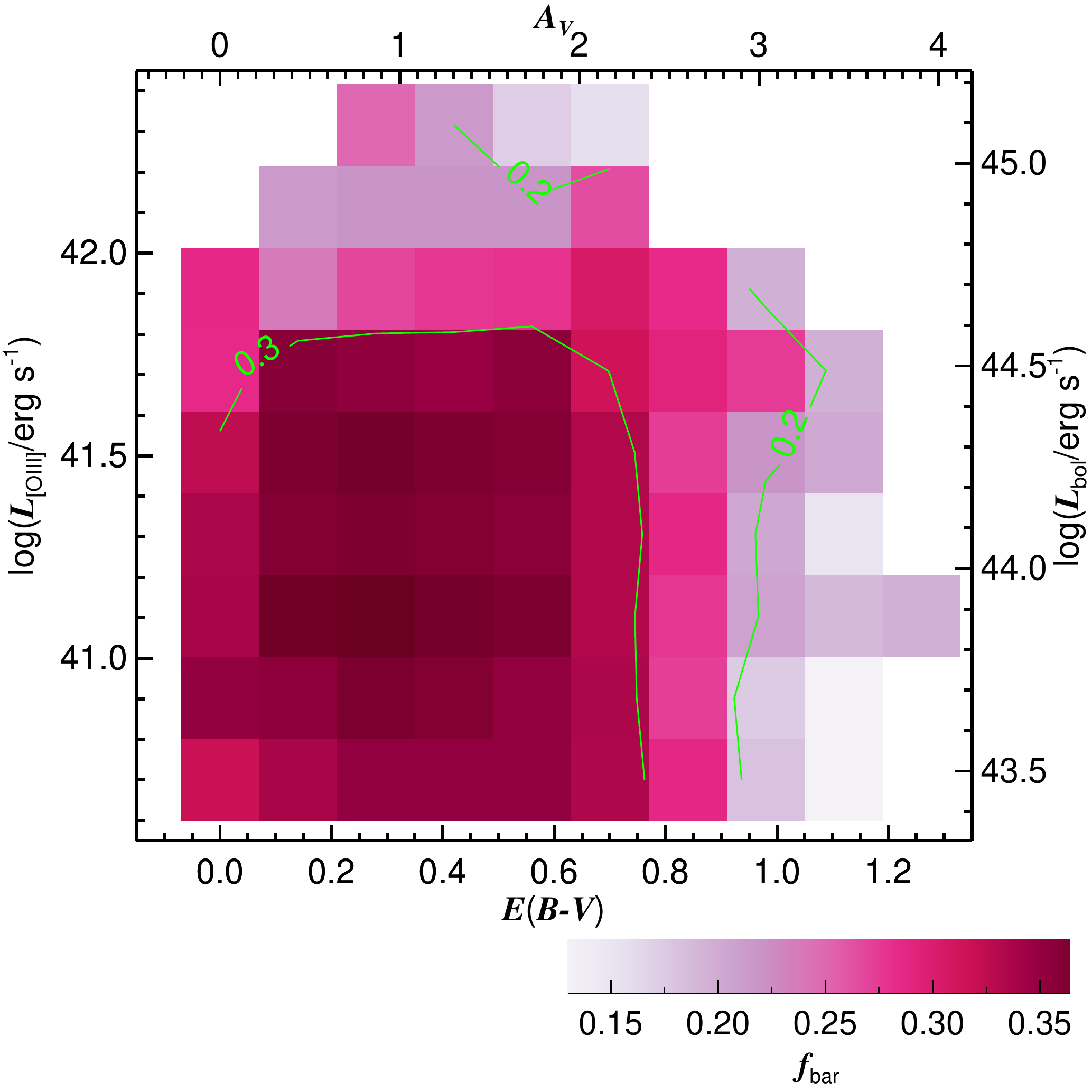}
\centering
\caption{Fraction of type 2 AGN hosts with bar structures ($f_\mathrm{bar}$) shown by colors and contours in the $E(B-V)$ versus $\log L_\mathrm{[O\,{\footnotesize III}]}$ plane. The color bar indicates the color-coded values of $f_\mathrm{bar}$, and the numbers in the middle of the contour lines represent the corresponding $f_\mathrm{bar}$ values. The method for generating this plot is similar to that used for Figure \ref{fig:2df}.
\label{fig:bar}}
\end{figure}

\subsection{Role of bar structures in merger-driven AGN triggering}\label{sec:bar}
Bar structures are common morphological features observed in galaxies in the local Universe \citep{Jogee2004,Lee2012a,Yoon2019,Yoon2020}. Some studies suggest that bars can play an important role in the triggering of AGN activity \citep{Galloway2015,Alonso2018}, although others report no significant connection between bars and AGNs \citep{Lee2012b,Cheung2015,Cisternas2015}. Therefore, it is useful to briefly examine how bar structures are connected to our results in order to clarify the role of bars in merger-driven AGN triggering, which may help provide a more complete understanding of AGN triggering.

For this examination, we identify bar structures in the AGN host galaxies using the visual inspection method described in Section \ref{sec:tidal}. We find that $32.4\%$ of AGN hosts in the full sample exhibit bar structures (242/748). This bar fraction ($f_\mathrm{bar}$) is consistent with the $\sim30\%$ reported for galaxies at low redshift in other studies \citep{Lee2012a,Yoon2019}. We cross-matched our bar classifications with the catalog of \citet{Dominguez2018}---which contains information on the probability of the presence of bar structures for $\sim670,000$ galaxies in the SDSS---in order to assess the reliability of our bar classifications. The morphological classifications in this catalog, including the possibility of the presence of bars, were obtained using deep-learning algorithms based on convolution neural networks, with separate models trained on the sample from Galaxy Zoo 2 \citep{Willett2013} and the catalog of \citet{Nair2010}. In examining 669 cross-matched AGN host galaxies, we find that $\gtrsim80\%$ of the bar classifications in the catalog are consistent with ours when galaxies with bar probabilities greater than $\sim50\%$--$60\%$ are classified as barred. For example, when galaxies with the bar probability PbarGZ2$>60\%$\footnote{Bar probability obtained from a model trained on the Galaxy Zoo 2 sample.} are defined as barred, $81.6\%$ of those in the catalog were also classified as barred by us, while $80.1\%$ of the non-barred galaxies in the catalog are likewise classified as non-barred following our visual inspection. Using a threshold of the bar probability PbarNair10$>60\%$\footnote{Bar probability derived from a model trained on the \citet{Nair2010} catalog.} to define barred galaxies, $85.6\%$ of the barred galaxies in the catalog were also classified as barred by us, whereas $79.4\%$ of the non-barred galaxies in the catalog were also classified as non-barred via our visual inspection. The $\gtrsim80\%$ agreement rate implies a potential systematic uncertainty in the bar fraction of up to $\sim0.04$-$0.07$ depending on the number of galaxies in each bin. Such a level of uncertainty is unlikely to invalidate our conclusion below.

We find that $f_\mathrm{bar}$  is $0.14\pm0.03$ for type 2 AGN hosts with tidal features, whereas $f_\mathrm{bar}$ for those without tidal features is $0.37\pm0.02$, indicating that AGN hosts with tidal features exhibit a significantly lower bar fraction. Likewise, barred AGN hosts exhibit a substantially lower tidal-feature fraction, $f_T=0.10\pm0.02$, than non-barred AGN hosts, for which $f_T=0.37\pm0.02$. Figure \ref{fig:bar} illustrates how the result for the bar fraction is connected to our overall results. The figure shows that $f_\mathrm{bar}$ is lower for AGNs with higher luminosities or greater dust obscuration, which exhibit relatively higher fractions of tidal features in our results. Specifically, AGNs with $\log L_\mathrm{[O\,{\footnotesize III}]}\ge41.9$ or $E(B-V)\ge0.7$ have $f_\mathrm{bar}=0.19\pm0.04$, whereas the other AGNs with $\log L_\mathrm{[O\,{\footnotesize III}]}<41.9$ and $E(B-V)<0.7$ exhibit $f_\mathrm{bar}=0.34\pm0.02$. 

These results imply that galaxy mergers or gravitational interactions capable of triggering AGNs may be unfavorable for the formation of bar structures, at least immediately after the merger events. This is also supported by other studies \citep{Lee2012a,Casteels2013,Ghosh2021}. In other words, bar structures do not appear to play a significant role in the merger-driven AGN-triggering mechanism, at least during the early post-merger phase.
However, if the results presented in Section \ref{sec:results} are interpreted within an evolutionary framework in which AGNs that are simultaneously obscured and luminous are at stages temporally closer to merger events than AGNs with lower luminosities and less dust obscuration, the combined results of Figures \ref{fig:2df} and \ref{fig:bar} suggest that bars may develop later as the merger remnants evolve into a relaxed stage and the tidal features gradually fade. Nevertheless, the theoretical picture remains incomplete. While bar formation in isolated disks and tidally induced bar growth have been studied extensively \citep{Hohl1971,Barnes1992,Berentzen2004,Peirani2009,Moetazedian2017}, the conditions under which bars arise in merger remnants are still not well established.

Alternatively, if our results presented in Section \ref{sec:results} are interpreted to suggest that galaxy mergers preferentially trigger AGNs that are simultaneously luminous and dust-obscured, bars may be relevant to the triggering of a nontrivial fraction of AGNs with lower luminosities and less dust obscuration, as implied by the high bar fraction observed in the AGNs in Figure \ref{fig:bar}.

We note that our bar classification is based on visual inspection, so the measured bar fraction should be regarded as a conservative estimate, and the true bar fraction could be somewhat higher if a more sophisticated bar-detection method were applied. Therefore, although the bar fraction of $\sim0.34$ in the low-$E(B-V)$, low-$L_\mathrm{[O\,{\footnotesize III}]}$ regime is broadly comparable to that reported for local galaxies in general and thus does not by itself imply a bar excess, it is nevertheless not negligible; this suggests that bars can play an important role in this part of parameter space. This, in turn, together with the above discussion, implies the importance of considering AGNs over a broad range of properties when investigating AGN triggering by bars and galaxy mergers.

One potential limitation in this bar-fraction analysis is that bars may be missed in optical images in the case of heavily obscured AGN hosts. Therefore, high-resolution, deep near-infrared imaging is required to investigate this issue more thoroughly, and this will be pursued in a future work.
\\

\section{Summary}\label{sec:summary}
Using a large sample of 748 type 2 AGNs at $z<0.063$, we examined the fraction of AGN hosts exhibiting merger features as a function of Balmer-decrement-derived $E(B-V)$ and internal-extinction-corrected $L_\mathrm{[O\,{\footnotesize III}]}$. This approach allowed us to disentangle the contributions of two key variables---intrinsic AGN luminosity and dust extinction---to the merger--AGN connection, thereby obtaining a more comprehensive understanding not only of this connection, but also of the evolutionary scenario of merger-triggered AGNs. 

Type 2 AGNs were selected using the BPT diagram from the MPA-JHU catalog, which is based on SDSS Data Release 8. The $E(B-V)$ values derived from the Balmer decrement and those from stellar continuum fitting are well correlated, while the Balmer decrement-based $E(B-V)$ is, on average, 0.35 higher. This suggests that the dust extinction affecting AGN emission lines is likely associated with dust on the scale of the host galaxy, and the line-of-sight extinction toward the narrow-line region is higher than the stellar light-weighted extinction of the central regions covered by the fiber aperture. As direct evidence of recent mergers, we used tidal features identified through a visual inspection of deep images from the DESI Legacy Imaging Survey.

Our main finding is that the fraction of type 2 AGN hosts with tidal features ($f_T$) is only significantly higher for AGNs that are simultaneously luminous and heavily dust-obscured. Specifically, AGNs with $\log L_\mathrm{[O\,{\footnotesize III}]}\gtrsim41.5$ and $E(B-V)\gtrsim0.7$ exhibit a high $f_T$ of $\sim0.7$. By contrast, AGNs with either low luminosity ($\log L_\mathrm{[O\,{\footnotesize III}]}\lesssim41.0$) or low dust obscuration ($E(B-V)\lesssim0.3$) show a low $f_T$ of $\lesssim0.2$.

This suggests that galaxy mergers preferentially trigger AGNs that are simultaneously luminous and dust-obscured, whereas high-luminosity type 2 AGNs without dust obscuration and dust-obscured AGNs with low luminosities are not strongly associated with merger-driven triggering. If we posit several assumptions---for example, that the AGN-activity timescale is comparable to that of tidal features---our result can be interpreted, despite certain caveats, within the framework of a merger-initiated evolution model for AGNs, suggesting that AGNs that are simultaneously obscured and luminous are temporally closer to merger events than those with lower luminosities and less dust obscuration. Our result also highlights the importance of AGN sample-selection methods and the ranges of AGN properties as critical factors in AGN studies.

We also investigated the connection between bar structures and our results in order to clarify the role of bars in merger-driven AGN triggering, using visually identified bars in our type 2 AGN sample. We find that bar structures do not appear to play a significant role in the merger-driven triggering mechanism, at least during the early post-merger phase. However, bars may emerge later as the merger remnants evolve into a relaxed stage and the tidal features fade. Another possibility is that they are relevant to the triggering of a nontrivial fraction of low-luminosity, less-dust-obscured AGNs.
\\

\section{Data availability}
Table \ref{table} is available at the CDS via \url{http://cdsweb.u-strasbg.fr/cgi-bin/qcat?J/A+A/}.
\\

\begin{acknowledgements}
This research was supported by Kyungpook National University Research Fund, 2025.
This research was supported by the National Research Foundation of Korea (NRF) grant funded by the Korea government (MSIT) (RS-2025-16064514).
Y. K. was supported by the faculty research fund of Sejong University in 2025 and the National Research Foundation of Korea (NRF) grant funded by the Korean government (MSIT) (2021R1C1C2091550 and RS-2026-25476464).
D.K. acknowledges the support by the National Research Foundation of Korea (NRF) grant (2021R1C1C1013580  and RS-2026-25491376) funded by the Korean government (MSIT).
\end{acknowledgements}

\end{document}